\documentclass[superscriptaddress,showpacs,showkeys,preprintnumbers,nofootinbib,amsmath,amssymb,aps,onecolumn,prd,floatfix]{revtex4-2}

 \newcommand{\be}[1]{\begin{equation}\label{#1}}
 \newcommand{\ba}[1]{\begin{eqnarray}\label{#1}}
 \newcommand{\ep}[1]{\epsilon_{#1}}

 \newcommand{\de}[1]{\delta_{#1}}

 \newcommand{\rd}{{\rm d}}
 
 \newcommand{\re}{{\rm e}}
 \newcommand{\pa}[1]{\left(#1\right)}
 \newcommand{\paq}[1]{\left[#1\right]}
 \newcommand{\pag}[1]{\left\{#1\right\}}

 \newcommand{\M}{{\rm M_{\rm P}}}
 
 \newcommand{\R}{{\mathcal R}}

 \def\ee{\end{equation}}
 \def\ea{\end{eqnarray}}

\usepackage[T1]{fontenc}
\usepackage[utf8]{inputenc}
\usepackage{tikz}

\begin{document}

\title{Superpotential method and the amplification of inflationary perturbations}
\author{Alexander~Yu.~Kamenshchik}
\email{kamenshchik@bo.infn.it}
\affiliation{Dipartimento di Fisica e Astronomia, Universit\`a di Bologna\\ and INFN,  Via Irnerio 46, 40126 Bologna,
Italy}
\author{Ekaterina~O.~Pozdeeva}
\email{pozdeeva@www-hep.sinp.msu.ru}
\affiliation{Skobeltsyn Institute of Nuclear Physics, Lomonosov Moscow State University,\\
 Leninskie Gory 1, 119991, Moscow, Russia}
 \author{Augustin Tribolet}
 \email{augustin.tribolet@outlook.com}
\affiliation{Dipartimento di Fisica e Astronomia, Universit\`a di Bologna\\ and INFN,  Via Irnerio 46, 40126 Bologna,
Italy}
\affiliation{Institut d'Astrophysique et de Géophysique, University of Liège, Allée du 6 août 19C, 4000 Liège, Belgium}
\author{Alessandro~Tronconi}
\email{tronconi@bo.infn.it}
\affiliation{Dipartimento di Fisica e Astronomia, Universit\`a di Bologna\\ and INFN, Via Irnerio 46, 40126 Bologna,
Italy}
\author{Giovanni~Venturi}
\email{giovanni.venturi@bo.infn.it}
\affiliation{Dipartimento di Fisica e Astronomia, Universit\`a di Bologna\\ and INFN,  Via Irnerio 46, 40126 Bologna,
Italy}
\author{Sergey~Yu.~Vernov}
\email{svernov@theory.sinp.msu.ru}
\affiliation{Skobeltsyn Institute of Nuclear Physics, Lomonosov Moscow State University,\\
 Leninskie Gory 1, 119991, Moscow, Russia}

\begin{abstract}
The superpotential method is a reconstruction technique which has proven useful to build exact cosmological solutions. We here employ the superpotential method in order to reconstruct the features necessary for the inflaton potential to drive inflation and lead to the amplification of the curvature perturbations. Such an amplification, at wavelengths shorter than those observed in the cosmic microwave background, is necessary in order to have a significant formation of primordial black holes after inflation ends. The technique is applied to the cases of a minimally coupled inflaton, to the nonminimal coupling case and to $f(R)$ theories of gravity. For such theories, a model dependent analysis of the features leading to the scalar spectrum enhancement is also presented.
\end{abstract}




\maketitle

\section{Introduction}
The hypothesis that black holes of primordial origin (PBHs) may constitute a relevant fraction of the energy density in our Universe was introduced more than 50 years ago in two seminal papers~\cite{Zeldovich:1967lct,Hawking:1971ei}. Moreover, the direct detections of gravitational waves (GW) by Ligo/Virgo~\cite{LIGOScientific:2016aoc} also attracts much attention to PBHs~\cite{Dolgov:2017aec,Carr:2020xqk,Ozsoy:2023ryl,Carr:2023tpt}.

The observed GWs originate from the coalescence of astrophysical objects with masses incompatible with our present knowledge of stellar collapse and are therefore, perhaps, of primordial origin~\cite{Dolgov:2017aec,Carr:2023tpt}. Recently, the possible existence of a large number of PBHs has therefore been reconsidered and related to the dark matter content~\cite{Carr:2020xqk,Carr:2023tpt,Espinosa:2017sgp}.

PBHs may be formed from the gravitational collapse of primordial density inhomogeneities generated during inflation~\cite{Dolgov:1992pu,Ivanov:1994pa,Garcia-Bellido:1996mdl,Karam:2022nym,Kristiano:2022maq} (see also the review~\cite{Ozsoy:2023ryl}). Their abundance and their mass can then be related to the features of the inflationary spectrum of curvature perturbations. A large amplification of the amplitude of the inflationary spectrum at scales shorter than those probed by the cosmic microwave background (CMB) is necessary in order for perturbations to collapse and form PBHs during the radiation domination era. A very finely tuned inflationary potential is needed in order to achieve the desired amplification. This tuning imposes quite severe constraints on the inflaton evolution and, from a technical point of view, building models of inflation leading to the desired scalar spectrum may be complicated. The conditions for PBH formation can be created both in one-field and in two-field inflationary models~\cite{Ketov:2021fww}. In this paper, we consider one-field models.

Scalar perturbations may then seed gravitational waves. This effect is tiny at the scales probed by the cosmic microwave background (CMB), whose amplitude is very small ($\sim 10^{-9}$), but may be relevant if some region of the inflationary spectrum is amplified by several orders of magnitude. Thus, PBHs formation may be accompanied by a stochastic GW background within a  frequency range of $\sim {\rm nHz}$ and related to the mass of the forming PBHs. A stochastic background of GWs in that range has been recently observed by NANOGrav and other pulsar timing array (PTA) collaborations. In the next few years increasingly precise estimates of the spectrum of such GWs will possibly shed light on the inflationary era at scales nowadays not accessible.

The scope of this article is to analytically reconstruct the inflationary evolution leading to an amplification of the curvature perturbations. The reconstruction procedure employs, whenever possible, the superpotential technique~\cite{Muslimov:1990be,Salopek:1990jq,Arefeva:2004odl,Townsend:2007aw} and is applied to a set of modified gravity theories~\cite{Kamenshchik:2012pw,Vernov:2019ubo}. The aim of this paper is not to build a complete inflationary potential but to find the general analytic features of inflationary models leading to an amplification. The corresponding spectrum of scalar perturbations is then estimated.

The article is organized as follows: in Sec II, we discuss the general features needed in order to obtain amplification. In Sec. III, we introduce the superpotential method, which is then applied to single field models with the inflaton minimally and nonminimally coupled to gravity.
The case of minimally  coupled inflaton is considered in Sec. IV. In Sec. V, we investigate models with nonminimal coupling, in particular induced gravity models.
In Secs. VI and VII, we discuss $f(R)$ theories and finally, in Sec. VIII, we illustrate our conclusions.

\section{Amplification of the spectrum}
Let us consider single-field minimally coupled models of inflation with $\phi(t)$ being the homogeneous inflaton.
In the spatially flat Friedmann--Robertson--Walker (FRW) metric with the interval
\be{metric}
\rd s^2=\rd t^2-a^2(t)\rd\vec x^2,
\ee
where  $a(t)$ is the  scale factor, the homogeneous Klein--Gordon equation for the inflaton has the following form:
\be{stdKG}
\ddot{\phi}+3H\dot{\phi}+\frac{\rd V}{\rd\phi}=0\,,
\ee
where the dot denotes differentiation with respect to the cosmic time $t$ and $H$ is the Hubble parameter $H\equiv \dot a/a$.

The constant-roll regime (CR) is a particular inflationary evolution, wherein slow-roll (SR) conditions are violated~\cite{Motohashi:2014ppa,Nojiri:2017qvx,Motohashi:2017vdc,Yi:2017mxs}.
In such a regime, one still has $\ep{1}=-\dot{H}/H^2<1$ corresponding to an accelerated expansion of the Universe, but in contrast with the SR regime, the resulting inflationary spectra can be an increasing function of the wave number $k$. This normally does not occur for the SR inflation where the resulting spectral amplitude for any given mode $k$ is proportional to $H^2$ evaluated at the time of the horizon crossing $k/(aH)\sim 1$. Since the Hubble parameter $H$ decreases in time, the amplitude is a (slowly) decreasing function of $k$.

If constant-roll solutions of the homogeneous Klein--Gordon equation exist, then they satisfy
\be{CR}
\ddot \phi-\beta H\dot \phi=0
\ee
where $\beta$ is a constant, or equivalently, the Hubble parameter satisfies the following equation:
\be{CRH}
\ddot H-2\beta H\dot H=0\,.
\ee

The existence of such solutions depends on the inflaton potential, whose particular form can be reconstructed from the comparison of Eqs.~(\ref{stdKG}) and (\ref{CR}) (see details in Refs.~\cite{Motohashi:2014ppa,Nojiri:2017qvx,Motohashi:2017vdc,Yi:2017mxs}).
Slow-roll (SR) parameters' hierarchy is used in the context of inflation as a replacement of the homogeneous degrees of freedom $H$ and $\phi$ (inflaton). The time variation of such parameters divided by $H$ is negligible if they are much smaller than unity.  Therefore, the adoption of the SR parameters instead of $H$, $\phi$ and their derivatives is useful to obtain an approximate form of the homogeneous dynamical equations. Moreover, the spectral indices of the inflationary spectra can be expressed in terms of the SR parameter. Throughout this article, we shall adopt two different hierarchies of the SR parameters. The Hubble flow function hierarchy is recursively defined as
\be{Hff}
\ep{0}=\frac{H_0}{H},\; \ep{i+1}=\frac{\dot{\ep{i}}}{H\ep{i}}
\ee
and the scalar field flow function hierarchy is
\be{Sff}
\de{0}=\frac{\phi}{\phi_0}, \;\de{i+1}=\frac{\dot{\de{i}}}{H\de{i}}\,,
\ee
where $H_0$ and $\phi_0$ are arbitrary constants.

The CR equation (\ref{CRH}) may be rewritten in terms of the SR parameters as
\be{USRsr}
2\ep{1}^2-\ep{1}\ep{2}+2\beta \ep{1}=0\stackrel{\ep{1}\neq 0}{\longrightarrow}2\ep{1}-\ep{2}+2\beta=0,
\ee
or equivalently
\be{USRsr2}
\de{1}\pa{\de{2}+\de{1}-\ep{1}}-\beta \de{1}=0\stackrel{\de{1}\neq 0}{\longrightarrow}\de{2}+\de{1}-\ep{1}-\beta=0,
\ee
and has a constant de Sitter solution ($\ep{1}=0$) and a time-dependent solution with $\ep{1}\rightarrow 0$ and $\ep{2}\rightarrow -2\beta$ in the $t\rightarrow \infty$ limit. Let us note that the second solution describes the evolution approaching the de Sitter solution and the $-2\beta$ limit for $\ep{2}$ is a consequence of the presence of the de Sitter solution. Indeed (as we already discussed in~\cite{Chataignier:2023ago}) if an inflaton-gravity system approaches a different attractor with $\ep{1}\rightarrow \ep{1,\infty}\neq 0$, then from the definition of $\ep{2}$ in (\ref{Hff}), one has $\ep{2}\rightarrow 0$. The existence of a de Sitter solution is a necessary but not sufficient condition in order to have such a nontrivial limit for $\ep{2}$. Similarly, one needs $\de{1}\rightarrow 0$ in order to have $\de{2}\rightarrow \de{2,\infty}\neq 0$.

Differentiating (\ref{USRsr}), one finds
\be{USRsr1}
2\ep{1}-\ep{3}=0,\;{\rm and}\quad 2\ep{1}\ep{2}-\ep{3}\ep{4}=0\Rightarrow \ep{2}=\ep{4}
\ee
and then $\ep{2i+1}=2\ep{1}$, $\ep{2i}=\ep{2}$ with $i$ a positive integer.

The existence of an ultraslow-roll (USR) phase during inflation is frequently considered in order to achieve the desired amplification of the inflationary perturbations. Such a phase may occur when the minimally coupled inflaton rolls close to an inflection point of its potential ($V_{,\phi}=V_{,\phi\phi}=0$) where its dynamics can be described by (\ref{CR}) with $\beta=-3$.


\section{Evolution Equations and the Superpotential Method}
\subsection{Evolution equations}
 Our overview will be general enough to include nonminimally coupled models; however, through a frame transformation we can always recast our model in the form of a minimally coupled inflaton-gravity system. Let us note that the superpotential method is similar to the Hamilton--Jacobi approach~\cite{Salopek:1990jq,Liddle:1994dx,Kinney:1997ne} since it uses the (homogeneous) scalar field values to parametrize the evolution. Indeed, it can be applied to cosmological models in which the evolution of the scalar field is monotonic. $f(R)$ gravity inflationary models will also be discussed in this article.

$f(R)$ gravity inflationary models~\cite{Starobinsky:1980te,Maeda:1987xf,Berkin:1990nu,Huang:2013hsb,Motohashi:2014tra,Miranda:2017juz,Ketov:2019toi,Ivanov:2021chn,Frolovsky:2022ewg,Pozdeeva:2022lcj,Brinkmann:2023eph,Saburov:2023buy,Saburov:2024und} as well as inflationary models with nonminimally coupled scalar fields~\cite{Spokoiny:1984bd,Accetta:1985du,Futamase:1987ua,Barvinsky:1994hx,Cervantes-Cota:1995ehs,Bezrukov:2007ep,Barvinsky:2008ia,Cerioni:2009kn,Ferrara:2010yw,Elizalde:2014xva,Elizalde:2015nya} are actively studied.
Generally, the models considered can be described by the following action:
\begin{equation}
\label{action}
S=\int d^4 x \sqrt{-g}\left[ U(\sigma)R-\frac{c}{2}g^{\mu\nu}\sigma_{,\mu}\sigma_{,\nu}+V(\sigma)\right],
\end{equation}
where $U(\sigma)$ and $V(\sigma)$ are differentiable functions of the scalar field $\sigma$, $g$ is the determinant of the metric tensor $g_{\mu\nu}$. We shall only consider the case with $U(\sigma)>0$. The case with $c=0$ corresponds to $f(R)$ gravity, whereas the case with $c=1$ corresponds to a standard scalar field.
Indeed (see ~\cite{Ivanov:2021chn}), $f(R)$ gravity models with
\be{actfR}
S_R=\int \rd^4x\sqrt{-g}f(R).
\ee
are equivalent to the following scalar-tensor gravity models
\begin{equation}
\label{actfRJF}
S_J=\int d^4 x \sqrt{-g} \left[ f_{,\sigma} (R-\sigma)+ f\right]~,
\end{equation}
where $\sigma$ is a scalar field without kinetic term.

In Secs. III C, IV and V, we shall consider the case of a standard scalar field and $c=1$. The $f(R)$ gravity models will be considered in Secs VI and VII.

For the evolution of a homogeneous scalar field on a spatially
flat FRW universe with the metric (\ref{metric}),
the Einstein equations derived from the action (\ref{action}) are as follows:
\begin{equation}
\label{equ00}
6UH^2+6\dot U H=\frac{c}{2}{\dot{\sigma}}^2+V,
\end{equation}
\begin{equation}
\label{equ11}
2U\left(2\dot H+3H^2\right)+4\dot U H+2\ddot U +\frac{c}{2}{\dot{\sigma}}^2-V=0,
\end{equation}
and the variation of (\ref{action}) with respect to $\sigma$ gives
the Klein--Gordon equation,
\begin{equation}
\label{dequsigma}
c\pa{\ddot \sigma+3H\dot\sigma}+V_{,\sigma}=6\left(\dot H +2H^2\right)U_{,\sigma}\,,
\end{equation}
where the subscript $ _{,\sigma}$ indicates the derivative with respect to the scalar field
$\sigma$. For a constant $U$, $c=1$ and $\sigma\rightarrow \phi$, Eq.~(\ref{dequsigma}) reduces to (\ref{stdKG}).
Combining  Eqs.~(\ref{equ00}) and (\ref{equ11}), one obtains
\begin{equation}
\label{equ01}
4U\dot H-2HU_{,\sigma}\dot{\sigma}+2\left(U_{,\sigma\sigma}{\dot{\sigma}}^2+U_{,\sigma}\ddot{\sigma}\right) +c{\dot\sigma}^2=0.
\end{equation}

\subsection{Evolution of the inflationary spectra}
In inflationary theories described by (\ref{action}), the Mukhanov--Sasaki equation for scalar inflationary perturbations has the form
\be{MS}
v''_k+\pa{k^2-\frac{z''}{z}}v_k=0
\ee
where the prime denotes the derivative with respect to conformal time and
\be{defz}
z\equiv \frac{\sqrt{c+3\frac{\dot U^2}{\dot \sigma^2 U}}}{1+\frac{\dot U}{2HU}}\frac{a\dot \sigma}{H}.
\ee
Correspondingly, the comoving curvature perturbation ${\mathcal R}_k$ is related to $v_k$ by
\be{R_kdef}
{\mathcal R}_k=\frac{v_k}{z}
\ee
and satisfies the following second-order differential equation
\be{eqRz}
\zeta^2\frac{\rd^2 \R_k}{\rd \zeta^2}+\paq{\frac{\ep{1}\ep{2}-2\pa{1-\ep{1}}\frac{\rd \ln z}{\rd N}}{\pa{1-\ep{1}}^2}}\zeta \frac{\rd \R_k}{\rd \zeta}+\frac{\zeta^2}{\pa{1-\ep{1}}^2}\R_k=0 \ ,
\ee
with $\zeta\equiv k/(aH)$ and the e-folding number $N\equiv \ln(a/a_0)$. In the long wavelength limit $\zeta\rightarrow 0$, Eq.~(\ref{eqRz}) admits a constant solution and, for constant SR parameters, a solution of the form $\zeta^B$ with
\be{Bsol}
B=1-\frac{\ep{1}\ep{2}-2\pa{1-\ep{1}}\frac{\rd \ln z}{\rd N}}{\pa{1-\ep{1}}^2}.
\ee
This second solution evolves as ${\mathcal R}_k\simeq {\rm e}^{-B(1-\epsilon_1)N}$ and increases or decreases depending on the sign of
\be{Phi}
\Phi\equiv B\pa{1-\ep{1}}=\frac{\pa{1-\ep{1}}^2-\ep{1}\ep{2}+2\pa{1-\ep{1}}\frac{\rd \ln z}{\rd N}}{\pa{1-\ep{1}}}.
\ee
The existence of an increasing solution ($\Phi<0$) is a sufficient condition for the amplification of the inflationary spectra. Moreover if $\Phi>0$ a positive spectral index $n_s-1$ signals an amplification. In such a case and for the general action (\ref{action}), one has
\be{nsm1gen}
n_s-1=2-\paq{\frac{\ep{1}\ep{2}}{(1-\ep{1})^2}-1}-\sqrt{\paq{\frac{\ep{1}\ep{2}}{(1-\ep{1})^2}-1}^2+4\frac{z_{,NN}+(1-\ep{1})z_{,N}}{z (1-\ep{1})^2}}
\ee
where the comma indicates differentiation. Let us note that $z$ depends on the choice of the inflationary action (\ref{action}) and $z_{,N}/z$, $z_{,NN}/z$ are model-dependent functions of the SR parameters $\ep{i}$'s and $\delta_{i}$'s.\\
For the inflationary models described by the action (\ref{action}) and in the limit $\ep{2i+1}\rightarrow 0$ and $\de{2i+1}\rightarrow 0$ one finds
\be{Phigen}
\Phi=3+2\de{2}\quad {\rm and}\quad n_s-1=3-\sqrt{\Phi^2}=-2\de{2}
\ee
and the relation between $\de{2}$ and the other SR parameters depends on the specific model considered.

\subsection{Superpotential method for models with nonminimal coupling}

In this subsection, we describe the superpotential method in its variant proposed for models with a nonminimally coupled scalar field~\cite{Kamenshchik:2012pw} ($c=1$). In the context of inflation, the superpotential method has been proposed in Ref.~\cite{Salopek:1990jq} (see also Refs.~\cite{Liddle:1994dx,Binetruy:2014zya}).

If we introduce the function $G(\sigma)$ defined as
\begin{equation}
\label{equsigma}
\dot \sigma=G(\sigma)
\end{equation}
and $\sigma=\sigma(t)$ is monotonic, then by the chain rule one has
\be{difft}
\frac{\rd}{\rd t}=G(\sigma) \frac{\rd}{\rd \sigma}
\ee
and Eq. (\ref{dequsigma}) takes the form
\be{equa}
4UH_{,\sigma}+2(G_{,\sigma}-H)U_{,\sigma}+(2U_{,\sigma\sigma}+1)G=0.
\end{equation}
Equation (\ref{equa}) contains two unknown functions: $G(\sigma)$ and $H(\sigma)$. Given one of them, the second one can be obtained as the solution of a linear first order differential equation.
If $G$ is given, then $H(\sigma)$ can be obtained as
\begin{equation}
\label{Hsigma}
H(\sigma)={}-\left[\int\limits^{\sigma}\frac{2G_{,\tilde\sigma}U_{,\tilde\sigma}+(2U_{,\tilde\sigma\tilde\sigma}+1)G}{4U^{3/2}}
\,\rd\tilde \sigma+c_0\right]\!\sqrt{U(\sigma)},
\ee
where  $c_0$ is an integration constant. Similarly, on using some given $H(\sigma)$, we can find $G(\sigma)$ as:
\begin{equation}
\label{SolF}
    G(\sigma)=\left[ \int\limits^\sigma \frac{U_{,\tilde{\sigma}}H-2U H_{,\tilde{\sigma}}}{U_{,\tilde{\sigma}}}  e^{\Upsilon} \rd\tilde{\sigma}+\tilde{c}_0\right]e^{-\Upsilon(\sigma)},
\end{equation}
where
\begin{equation*}
    \Upsilon(\sigma)\equiv\frac{1}{2}\int\limits^{\sigma}\frac{2U_{,\tilde\sigma\tilde\sigma}+1}
    {U_{,\tilde\sigma}}\,d\tilde\sigma
\end{equation*}
and $\tilde c_{0}$ is an integration constant. Let us note that, in the minimally coupled case ($U={\rm const}$), Eq. (\ref{SolF}) no longer holds. Indeed, Eq. (\ref{equa}) becomes an algebraic equation, which fixes $G$ in terms of $H_{,\sigma}$. In such a case, for any function $H(\sigma)$ the reconstruction procedure can be analytically fulfilled, and the potential can be obtained as a function of $\sigma$.
Once the three functions $H$, $G$ and $U$ of $\sigma$ are known, the corresponding potential $V(\sigma)$ can be easily obtained by inverting the Friedmann equation (\ref{equ00}):
\begin{equation}
\label{potentialV}
V(\sigma)=6UH^2+6U_{,\sigma}GH-\frac{1}{2}G^2=6H^2U\pa{1+3\frac{U_{,\sigma}^2}{U}}-\frac{\pa{G-6U_{,\sigma}H}^2}{2}.
\end{equation}
In principle, the time evolution of $\sigma(t)$ and $H(t)$ can also be found by integrating Eq.~(\ref{equsigma}), but this goes beyond the scope of this paper.

\section{Minimally coupled inflaton}
\subsection{Inflationary model}
Let us first consider the minimally coupled inflaton case\footnote{We conventionally use $\phi$ to indicate the scalar field minimally coupled to gravity and $\sigma$ to indicate the nonminimally coupled inflaton.} with $U=U_0=\M^2/2$. 
For such a case, Eq.~(\ref{equ01})  reduces to the acceleration equation
\be{acceq1}
\dot{H}={ } -\frac{\dot\phi^2}{2\M^2}
\ee
and through Eqs.~(\ref{equsigma}) and (\ref{difft}) one finds
\be{acceq2}
G={ }-2\M^2 H_{,\phi}.
\ee
If $H(\phi)$ is fixed, then $G(\phi)$ can be simply obtained by differentiating $H$ as illustrated by (\ref{acceq2}). Therefore, one can straightforwardly build exact potentials starting from any given Hubble parameter expression $H(\phi)$.  The expression (\ref{potentialV}) simplifies and one thus obtains
\be{VGR}
V(\phi)=3\M^2 H^2-2\M^4 H_{,\phi}^2.
\ee
For example for
\be{HUSR}
H=h_0 \cosh\paq{A\pa{\frac{\phi}{\phi_0}-1}}
\ee
and $A^2=3\phi_0^2/(2\M^2)$ one recovers the constant potential and USR evolution.
The SR parameters can be calculated in terms of $H(\phi)$ and its derivatives~\cite{Liddle:1994dx}. From the definitions (\ref{Hff}) and (\ref{Sff}) one has
\be{SR1GR}
\ep{1}={}-\frac{\dot{ H}}{H^2}={}-G\frac{H_{,\phi}}{H^2}=2\M^2\frac{H_{,\phi}^2}{H^2}, \;\de{1}=\frac{\dot \phi}{H \phi}={}-2\M^2\frac{H_{,\phi}}{H\phi}
\ee
and
\be{SR2GR}
\ep{2}=\frac{\dot{\ep{1}}}{H\ep{1}}={}-4\M^2\frac{H_{,\phi\phi}}{H}+2\ep{1},\; \de{2}=\frac{\dot{\de{1}}}{H\de{1}}=\frac{\ep{2}}{2}-\de{1}\,,
\ee
where expressions (\ref{SR2GR}) are very similar to the constant-roll conditions (\ref{USRsr}) and (\ref{USRsr2}).
We are interested in solutions with the scalar field and the Hubble parameter asymptotically approaching some (de Sitter) fixed point: $\phi\rightarrow \phi_0$, $H\rightarrow H_0$.
These solutions are realized if
\be{attGR}
\lim_{\phi\rightarrow \phi_0}\dot \phi=0,\quad \Rightarrow\quad \lim_{\phi\rightarrow \phi_0} H_{,\phi}\rightarrow 0.
\ee
If $H(\phi)$ is regular in $\phi_0\neq 0$, then its Taylor expansion around $\phi_0$ has the following form:
\be{HserdS}
H(\phi)=\sum_{n=0}^{\infty}h_n \pa{\frac{\phi}{\phi_0}-1}^n\equiv \sum_{n=0}^{\infty}h_n \pa{\frac{\delta\phi}{\phi_0}}^n\,,
\ee
and the condition (\ref{attGR}) is satisfied if $h_1=0$. Correspondingly $\de{1},\ep{1}\rightarrow 0$ and, to the leading order, one has
\be{attGR2}
\dot \phi\simeq -4\M^2 \frac{h_2}{\phi_0}\left(\frac{\phi}{\phi_0}-1\right).
\ee

We observe from (\ref{attGR2}) that $\phi_0$ is an attractor if $h_2/\phi_0>0$. Let us note that $h_0>0$ is a necessary condition in order to have inflation close to the attractor. If such conditions are satisfied, then one can calculate the SR parameters (\ref{SR2GR}) in the $\phi\rightarrow \phi_0$ limit and obtain
\be{SR2GRatt}
\lim_{\phi\rightarrow \phi_0}\ep{2}={}-8\frac{\M^2}{\phi_0^2}\frac{h_2}{h_0},\quad\lim_{\phi\rightarrow \phi_0}\de{2}={}-4\frac{\M^2}{\phi_0^2}\frac{h_2}{h_0}\equiv{}-\gamma.
\ee
Let us note that $\ep{2}$ and $\de{2}$ are different from zero if $h_2\neq 0$. The coefficient $h_2$ must be positive in order for $\ep{1}$ and $\de{1}$ to be decreasing functions of time close to $\phi_0$. \\
The expressions (\ref{SR2GRatt}) seem to indicate that $\ep{2}$ and $\de{2}$ may take arbitrary values, depending on the choice of $h_2$ and $h_0$. However, on calculating the potential around the attractor $\phi_0$ and imposing the condition for the stability of the solutions, one finds that for consistency $\gamma>-6$ and correspondingly $-6<\ep{2}<0$.\\

On calculating the successive SR parameters and evaluating their value at $\phi_0$, one finds a succession of zero and constant values in analogy with the CR case, as illustrated in our preceding article \cite{Chataignier:2023ago}.\\
Indeed, on integrating (\ref{attGR2}), one finds
\be{d0}
\de{0}=\frac{\phi}{\phi_0}\simeq1+\re^{-4\M^2\phi_0^{-2} h_2t}\simeq 1+\re^{-\gamma N}\quad{\rm and}\quad H\sim h_0+h_2\re^{-2\gamma N}.
\ee

The succession of zero and constant values can be verified by explicitly calculating the $\de{i}$'s and the $\ep{i}$'s starting from (\ref{d0}).

Once the SR parameters close to the attractor are calculated, one can calculate the parameter $\Phi$ and the spectral index $n_s-1$ on using the general expressions (\ref{Phi}) and (\ref{nsm1gen}). One has
\be{PhiGR}
\Phi=3+\ep{2}=3-2\gamma
\ee
and, when $\gamma>3/2$, $\Phi$ is negative, signaling the presence of an increasing solution of the Mukhanov--Sasaki equation (and thus of an amplification similarly to what occurs for the USR regime). In contrast, if $\gamma<3/2$, then the increasing solution is not present and one can calculate the spectral index
\be{nsm1GR}
n_s-1={}-\ep{2}=2\gamma
\ee
and observe that a blue-tilted spectrum (increasing) is obtained for $0<\gamma<3/2$.

\subsection{Stability}
Let us now study the stability of the de Sitter solution ($\phi=\phi_0$) with respect to homogeneous and isotropic perturbations. On expanding the homogeneous Klein--Gordon equation (\ref{stdKG}) around the scalar field fixed point $\phi_0$, to linear order one finds
\be{kgexp}
3\M^2V_{,\phi}(\phi_0)+\pa{V(\phi_0)\delta \phi_{,NN}+3 V(\phi_0)\delta \phi_{,N}+3\M^2V_{,\phi\phi}(\phi_0)\delta\phi}=0
\ee
where the first term is zero, $\delta \phi=\phi-\phi_0$, $V(\phi_0)=3\M^2h_0^2$,
\be{exppotstab}
V_{,\phi\phi}(\phi_0)=12\frac{\M^2}{\phi_0^2}h_0h_2\pa{1-\frac{4}{3}\frac{h_2}{h_0}\frac{\M^2}{\phi_0^2}}=\frac{V(\phi_0)}{\M^2}\pa{\gamma-\frac{\gamma^2}{3}}.
\ee
and $h_0$ and $h_2$ are the Taylor coefficients of the expansion of $H$ around $\phi_0$. Let us note that $V(\phi_0)$ and $V_{,\phi\phi}(\phi_0)$ are invariant with respect to the substitution $h_2\rightarrow -h_2+3\phi_0^2h_0/(4\M^2)$, where $\phi_0^2$ is the scalar field attractor and the coefficient $3/(4\M^2)$ is $-1/2$ the ratio between the coefficient in front $H^2$ and that in front of $H_{,\phi}^2$ in (\ref{VGR}). This last result is quite general and will be also applied to the nonminimal coupling case. This invariance is simply related to the fact that for any given potential, on linearizing the KG equation close to $\phi_0$, one finds two independent solutions. Indeed, the linearized equation for $\delta\phi$ is easily solved by
\be{solkg}
\delta\phi=c_1\re^{-N\pa{3-\gamma}}+c_2\re^{-\gamma N}.
\ee
The solution (\ref{solkg}) is stable if
\be{dSstab1}
0<\gamma<3,
\ee
and this result coincides with the stability condition for constant-roll solutions.
If the condition (\ref{dSstab1}) is satisfied, $\delta\phi$ decreases in time for any choice of the initial conditions $c_1$ and $c_2$. The second exponential decreases slower than the first if $0<\gamma<3/2$ and dominates close to the attractor. It may also dominate, for a non-negligible amount of time, if $3/2<\gamma<3$ and $c_1/c_2\ll1$. In such a case, one essentially recovers the solution (\ref{d0}) and the expressions (\ref{PhiGR}) and (\ref{nsm1GR}).  The first exponential dominates over the second close to the attractor if $3/2<\gamma<3$ or, for a non-negligible period, if $0<\gamma<3/2$ and $c_2/c_1\ll 1$. In this case, one finds
\be{deS}
\de{1}\sim 0,\quad \de{2}\sim -3+\gamma \quad{\rm and}\quad \ep{2}\sim2\de{2}
\ee
and correspondingly
\be{PHInsm1}
\Phi={}-3+2\gamma\quad{\rm and}\quad n_s-1=6-2\gamma.
\ee

\subsection{Curvature perturbation and PBHs}
We conclude this section with some general observation regarding the formation of PBHs which is directly connected with our study of the amplification of curvature perturbations. The first estimate of the collapse threshold for PBH is dated back in the 1970s when Carr \cite{Carr:1975qj} evaluated such a threshold to be
\be{pbhth}
\delta_{k,\rm th}\equiv\frac{\delta \rho_k}{\rho}\simeq c_s^2\,,
\ee
where $c_s=1/\sqrt{3}$ during the radiation domination and the density contrast must be evaluated at the horizon reentry. Through the Poisson equation it is possible to relate $\delta_{k}$ and $\mathcal R_k$:
\be{poisson}
\delta_k\simeq -\frac{4}{9}\pa{\frac{k}{aH}}^2\mathcal R_k.
\ee
Besides the gauge invariant quantity $\mathcal{R}_k$ describing the curvature perturbation of comoving slices, one also often finds in the literature~\cite{Chen:2013kta,White,Celoria:2019oiu} a slightly different quantity, $\zeta_k$ to describe curvature in a gauge invariant way (it is the curvature of uniform density slices) and defined by
\be{defzeta}
\zeta_k\equiv \Psi-\frac{H}{\dot \rho}\delta \rho
\ee
where $\Psi$ is the spatial curvature perturbation. In GR and for single-field inflation the difference between these two gauge invariant quantities is
\be{diffcurv}
\zeta_k-\mathcal R_k=\frac{2\rho}{9(\rho+P)}\pa{\frac{k}{aH}}^2\Psi\,,
\ee
where the longitudinal gauge has been considered to simplify the expression. The denominator on the right-hand side of (\ref{diffcurv}) is $\dot \phi^2=\de{1}^2\phi^2H^2$ and, thus, it simplifies to
\be{diffcurv2}
\zeta_k-\mathcal R_k=\frac{\M^2}{3\de{1}^2\phi^2}\pa{\frac{k}{aH}}^2\Psi
\ee
which during SR and for modes outside the horizon is negligible, since $k/(aH)$ decreases as $\re^{-N}$. This is not always the case if more general inflationary models are considered (see, for example~\cite{Chen:2013kta,Celoria:2019oiu}). If CR evolution is considered one may argue that $\zeta_k$ and $\mathcal{R}_k$ differ since one has $\de{1}\sim \re^{\de{2}N}$, with $\de{2}<0$ and there is a (partial) compensation between $k/(aH)$ and $\de{1}^{-1}$.  In order to extract the behavior of $\zeta_k$ and compare it with $\mathcal R_k$ the equations for such quantities must be explicitly compared. On doing so for GR with a minimally coupled inflaton one finds that for CR background evolution also such quantities obey the same equation and therefore, are equivalent to studying the amplification of curvature perturbations.

Let us note that the same conclusion holds for single-field inflationary models with nonminimal coupling, described by action (\ref{action}). This is the consequence of the fact that both $\zeta_k$ and $\mathcal{R}_k$ are invariant on employing a transformation between the Einstein and the Jordan frames~\cite{White,Makino:1991sg,Weinberg:2003sw,Chiba:2008ia,Sugiyama:2010zz,Kubota:2011re}.

\section{Nonminimally coupled inflaton}
When the case with a nonminimal coupling is considered, the superpotential method can be applied but some of the above simplifying assumptions do not hold. Indeed, one cannot easily relate $G$ and $H_{,\sigma}$ and easily express the potential and the SR parameters in terms of a single function and its derivative analogously to what we did in the minimally coupled case. This occurs since (\ref{equa}) contains the first derivatives of both $G$ and $H$ and for any given $H(\sigma)$, then $G(\sigma)$ must be obtained by integration (not differentiation).
In such a case, it is more convenient to reformulate the superpotential method in terms of a different set of functions as suggested in ~\cite{Skugoreva:2014gka} (see also Ref.~\cite{Pozdeeva:2016cja}).
One may define an effective potential
\begin{equation}
\label{Veff}
V_{\rm eff}(\sigma)=\frac{U_0^2V(\sigma)}{U^2(\sigma)},
\end{equation}
the new, dimensionless, function
\begin{equation}
\label{A}
A=\frac{U_0^2}{U^2}\left(1+\frac{3U_{,\sigma}^2}{U}\right)
\end{equation}
and
\be{Ydef}
Y=\sqrt{\frac{U_0}{U}}\pa{H+\dot \sigma\frac{U_{,\sigma}}{2U}}\,.
\ee
 In terms of these functions, Eqs.~(\ref{equ00}) and (\ref{equ01}) take the following form:
\begin{equation}
\label{equP}
6U_0 Y^2=\frac{A}{2}{\dot\sigma}^2+V_{\rm eff},
\end{equation}
which has the same structure as the Friedmann equation (\ref{equ00}) for a minimally coupled inflaton. The corresponding acceleration equation is
\begin{equation}
\label{Fr21Qm}
\dot Y={}-\frac{A\sqrt{U}}{4U_0^{3/2}}\,{\dot\sigma}^2 \Leftrightarrow Y_{,\sigma}=-\frac{A\sqrt{U}}{4U_0^{3/2}}\,G
\end{equation}
with $G=\dot \sigma$.
Furthermore, the Klein--Gordon equation~(\ref{dequsigma}) takes the following form:
\begin{equation}\label{KGequVeff}
\ddot\sigma={}-3\sqrt{\frac{U}{U_0}}Y\dot{\sigma}-\frac{A_{,\sigma}}{2A}{\dot{\sigma}}^2-\frac{{V_{\rm eff}}_{,\sigma}}{A}.
\end{equation}
Therefore, the condition ${V_{\rm eff}}_{,\sigma}=0$ clearly corresponds to de Sitter solutions. It has been shown in Ref.~\cite{Skugoreva:2014gka} that such solutions are stable for ${V_{\rm eff}}_{,\sigma\sigma}>0$ and unstable for ${V_{\rm eff}}_{,\sigma\sigma}<0$, provided the condition $U(\sigma)>0$ is satisfied. Let us note that any nonminimally coupled model can be transformed to the corresponding EF model through a conformal transformation of the metric and a field redefinition. In the EF, the effective potential $V_{eff}$ is equal to the potential~$V$ and $Y(\sigma)$ is the Hubble parameter.

Once the functional dependence of $U(\sigma)$ and $Y(\sigma)$ is properly chosen, one easily gets the corresponding effective potential
\begin{equation}
\label{VeffW}
V_{\rm eff}(\sigma)=2U_0\left(3Y^2-\frac{4U_0^2}{A\,U}Y_{,\sigma}^2\right)
\end{equation}
and the associated inflaton potential
\begin{equation}
\label{VW}
V(\sigma)=2\frac{U^2}{U_0}\left(3Y^2-\frac{4U_0^2}{A\,U}Y_{,\sigma}^2\right)=6\frac{U^2}{U_0}Y^2-\frac{8U^4}{U_0\left(3U_{,\sigma}^2+U\right)}Y_{,\sigma}^2.
\end{equation}

Finally, one finds
from Eqs.~(\ref{Ydef}) and (\ref{Fr21Qm}) that
\be{JFEF}
H=\sqrt{\frac{U}{U_0}}\pa{Y+\frac{2U\,U_{,\sigma}Y_{,\sigma}}{U+3U_{,\sigma}^2}},\quad G={}-\frac{4U^{5/2}Y_{,\sigma}}{\sqrt{U_0}\left(U+3U_{,\sigma}^2\right)}
\ee
Thus, on following the same procedure as used for the minimally coupled case, one can find, for any given choice of $U(\sigma)$, the conditions for $Y$ leading to specific inflationary evolutions in the Jordan frame.
\subsection{Induced gravity models}
To illustrate the proposed method let us work out in detail the induced gravity (IG) case with
\begin{equation}\label{IG}
U=\frac{\xi\sigma^2}{2}.
 \end{equation}
The IG term naturally arises in an inflationary model inspired by particle theories, when quantum effects are taken into account~\cite{Spokoiny:1984bd,Accetta:1985du,Barvinsky:1994hx,Bezrukov:2007ep,Barvinsky:2008ia,Cerioni:2009kn,Elizalde:2014xva,Elizalde:2015nya}.
From Eq.~(\ref{JFEF}), one has
\be{FHig}
G={}-\alpha \sigma^3Y_{,\sigma},\qquad H=\beta \sigma Y+\alpha\sigma^2Y_{,\sigma}\,,
\ee
with $\beta\equiv \sqrt{\xi}/\M$ and $\alpha\equiv 2\xi \beta/(1+6\xi)$.
One can easily write the corresponding potential\footnote{
Using
$\sigma=\sigma_0\exp\left(\frac{\xi\phi}{\sqrt{2U_0(\xi+6\xi^2)}}\right)$ and Eq.~(\ref{Veff}),
one can calculate the potential in the EF in terms of $\phi$.}

\be{Vig2}
V=\frac{\xi^2 \sigma^4\paq{3(1+6\xi)Y^2-2\xi\sigma^2 Y_{,\sigma}^2}}{(1+6\xi)\M^2},
\ee
and calculate the SR parameters
\be{de1ig}
\de{1}\equiv \frac{\dot \sigma}{H\sigma}={}-\frac{\alpha \sigma Y_{,\sigma}}{\beta Y+\alpha \sigma Y_{,\sigma}}={}-\frac{2\xi\sigma\,Y_{,\sigma}}{2\xi\sigma\,Y_{,\sigma} +\left(6\xi+1\right) Y}
,
\ee
\be{de2ig}
\de{2}\equiv\frac{\dot{\de{1}}}{H\de{1}}={}-\frac{\beta\,Y\,\delta_1+\sigma\,Y_{,\sigma}\left(2\alpha(1+\delta_1)+\beta\delta_1\right)+\alpha(1+\delta_1)\sigma^2\,Y_{\sigma\sigma}}{\beta\,Y+\alpha\sigma\,Y_{,\sigma}}
\ee
in terms of $Y$ and its derivatives.
If the inflaton evolves to some fixed point $\sigma_0$ (without loss of generality we consider $\sigma_0>0$) and we Taylor expand $Y$ around it
\be{TexpY}
Y=\sum_{n=0}^{\infty} y_n\pa{\frac{\sigma}{\sigma_0}-1}^n\,.
\ee
One may then easily observe that $\de{1}(\sigma_0)=0$, provided the first two coefficients of (\ref{TexpY}) satisfy the conditions $y_0\neq0$ and $y_1=0$. Then one has $\de{2}=-1-2\alpha/\beta \pa{y_2/y_0}$ at $\sigma_0$. The conditions $\de{1}(\sigma_0)=0$, $\de{2}(\sigma_0)\neq0$ are CR-like conditions, similar to those discussed before Eq. (\ref{USRsr1}), which may be associated to an amplifying inflationary phase.

A very simple potential possibly leading to an amplification can be obtained by choosing $Y=y_0+y_2\pa{\sigma/\sigma_0-1}^2$. In such a case,
\be{sys}
\sigma Y_{,\sigma}=2y_2\frac{\sigma}{\sigma_0}\pa{\frac{\sigma}{\sigma_0}-1}
\ee
and the potential is
\be{expotig}
V=\frac{\xi^2\sigma^4y_2^2\pag{3(1+6\xi)\paq{\frac{y_0}{y_2}+\pa{\frac{\sigma}{\sigma_0}-1}^2}^2-8\xi \frac{\sigma^2}{\sigma_0^2}\pa{\frac{\sigma}{\sigma_0}-1}^2}}{\M^2(1+6\xi)}.
\ee
For IG, the expression for $\Phi$, defined by Eq.~(\ref{Phi}), close to the de Sitter attractor ($\de{1}\rightarrow 0$) is
\be{PhiIG}
\Phi=3+2\de{2}=3-\frac{8\xi}{1+6\xi} \frac{y_2}{y_0}.
\ee
If $y_2/y_0>3(1+6\xi)/(8\xi)$, then $\Phi$ is negative and signals the presence of an increasing solution of the comoving curvature perturbation equation (\ref{eqRz}) which determines an amplification of the spectrum. In contrast, when $y_2/y_0<3(1+6\xi)/(8\xi)$, $\Phi$ is positive and the increasing solution is not present. The spectral index is given by (\ref{nsm1gen}) and one obtains:
\be{nsm1IG}
n_s-1=-2\de{2}=\frac{8\xi}{1+6\xi} \frac{y_2}{y_0}
\ee
and is then positive for $0<y_2/y_0<3(1+6\xi)/(8\xi)$. In this latter case one has the amplification.
Let us note that the discussion in Sec. III C (3.3) about solutions and stability can be straightforwardly applied to this case. In particular, concerning stability, it can be deduced from that of the corresponding solutions in the EF. Moreover, given the expression for the potential (\ref{Vig2}) and on substituting
\be{secondsolIG}
y_2\rightarrow -y_2+\frac{3\pa{1+6\xi}}{4\xi}y_0
\ee
one finds the second solutions close to $\sigma_0$. When such a solution dominates, one has
\be{Phinsm1IG}
\Phi=-3+\frac{8\xi}{1+6\xi} \frac{y_2}{y_0}
\quad{\rm and}\quad n_s-1=6-\frac{8\xi}{1+6\xi} \frac{y_2}{y_0}.
\ee

Let us also obtain the (general) behavior close to $\sigma_0\,$, where $Y\sim y_0+y_2\pa{\sigma/\sigma_0-1}^2$. One has
\be{Figlim}
\dot \sigma\simeq -2y_2\alpha \sigma_0^2\pa{\frac{\sigma}{\sigma_0}-1}, H\simeq \beta \sigma_0 y_0\paq{1+\pa{1+2\frac{\alpha  y_2}{\beta y_0}}\pa{\frac{\sigma}{\sigma_0}-1}}\,,
\ee
where we observe that $H$ now approaches the constant value linearly on varying the field $\sigma$ and $\de{0}$ has a ``time'' dependence similar to (\ref{d0}).

\section{$f(R)$ Theories}
\label{FRgravity}

The approach illustrated in previous sections can also be applied to the $f(R)$ theories, described by the action~\eqref{actfR}.

Considering flat FRW metric in absence of additional sources of matter, the dynamical equations reduce to the Friedmann-like equation
\be{frfR}
3 F H^2=\frac{1}{2}\pa{FR-f}-3H\dot F\,,
\ee
where the dot denotes the differentiation with respect to the cosmic time and $F(R)=\rd f/\rd R$.
For such theories, the Jordan to Einstein frame transformation consists of two steps. Starting from the action (\ref{actfR}) one rewrites it as the action~\eqref{action} with $c=0$.
Then by the conformal transformation
\be{conffR}
\tilde g_{\mu\nu}=\frac{2 F(\sigma)}{\M^2}g_{\mu\nu}
\ee
and the following redefinition of the scalar field
\be{sffR}
\phi=\sqrt{\frac{3}{2}}\M\ln\frac{2F(\sigma)}{\M^2}\,,
\ee
one obtains the EF action
\be{EFfR}
S_E=\int\rd^4 x\sqrt{-\tilde g}\paq{\frac{\M^2}{2}\tilde R-\frac{1}{2}\tilde g^{\mu\nu}\partial_\mu\phi\partial_\nu\phi+\tilde{V}}
\ee
where the tilde indicates physical quantities in the EF and $\phi$ is the minimally coupled ``inflaton''.
Let us note that the theories described by actions (\ref{actfR}) and (\ref{actfRJF}) are equivalent. If one calculates some observable in one of the two ``frames'' the result does not change. In contrast, the EF action \eqref{EFfR} is not equivalent to (\ref{actfR}). In particular from Eq.~\eqref{sffR}, it follows that $F(R)$ should be positive. In principle, one may proceed as illustrated for the nonminimal coupling case discussed in  the previous section. In this context, however, our aim is to search for the possible form of $f(R)$ leading to an amplification.

We note that once the potential or the Hubble parameter in the EF are calculated, the reconstruction of $f(R)$  is not straightforward~\cite{Ivanov:2021chn}. Indeed, the function $f(R)$ can be expressed only parametrically in terms of the field $\phi$ as follows~\cite{Motohashi:2017vdc}:
\be{Rparam}
R=\pa{\frac{\sqrt{6}\tilde{V}_{,\phi}}{\M}+\frac{4\tilde{V} }{\M^2}}\re^{\sqrt{\frac{2}{3}}\frac{\phi}{\M}}
\ee
and
\be{fparam}
f=\frac{\M^2}{2}\pa{\frac{\sqrt{6}\tilde{V}_{,\phi}}{\M}+\frac{2\tilde{V}}{\M^2}}\re^{2\sqrt{\frac{2}{3}}\frac{\phi}{\M}}\,.
\ee
Since the expression (\ref{Rparam}) cannot always be inverted, the analytic form of the $f(R)$ function can only be reconstructed in particular cases. The problem of the reconstruction of $f(R)$ models with exact solutions has already been studied in Ref.~\cite{Vernov:2019ubo}, where $f(R)$ was obtained for an EF potential of the form
\be{VfRef}
\tilde V=\frac{\M^2}{2}\pa{C_2\re^{-2\sqrt{\frac{2}{3\,}}\frac{\phi}{\M}}+C_1\re^{{} - \sqrt{\frac{2}{3\,}}\frac{\phi}{\M}}+C_{3}\re^{\omega
\sqrt{\frac{2}{3\,}}\frac{\phi}{\M}}},
\ee
where $C_i$ and $\omega$ are constants. This potential corresponds to
\be{fRab}
f(R)=\frac{\M^2}{2}\paq{C_3(\omega+1)\pa{\frac{R-C_1}{C_3(\omega+2)}}^\frac{\omega+2}{\omega+1}-C_2}.
\ee
On following a procedure similar to that illustrated for the IG case, we can calculate the Hubble flow function SR parameters {\color{red}(\ref{Hff})} in the JF in terms of the EF Hubble parameter $\tilde{H}(\phi)$ and its derivatives.
If $t$ is the cosmic time in the JF and $\tilde{t}$ is the cosmic time in the EF, then from Eqs.~(\ref{conffR}) and (\ref{sffR}) it follows that
\be{dtJdtE}
\frac{d\tilde t}{d t}=\re^{\frac{\phi}{\sqrt{6}\M}}
\ee
and one has
\be{HfR}
H=\re^{\frac{\phi}{\sqrt{6}\M}}\pa{\tilde{H}+\sqrt{\frac{2}{3}}\M \tilde{H}_{,\phi}}.
\ee
On further differentiating, one finds expressions for the JF slow-roll parameters:
\be{e1fR}
\ep{1}=2\M^2\tilde{H}_{,\phi}\frac{\frac{\tilde{H}}{\sqrt{6}\M}+\frac{4}{3}\tilde{H}_{,\phi}+\sqrt{\frac{2}{3}}\M \tilde{H}_{,\phi\phi}}{\pa{\tilde{H}+\sqrt{\frac{2}{3}}\M \tilde{H}_{,\phi}}^2}
\ee
and
\be{e2fR}
\begin{split}
\ep{2}&={}-\frac{2\M^2\tilde{H}_{,\phi}}{\tilde{H}+\sqrt{\frac{2}{3}}\M \tilde{H}_{,\phi}}\times\\
&\pa{\frac{\tilde{H}_{,\phi\phi}}{\tilde{H}_{,\phi}}-2\frac{\tilde{H}_{,\phi}+\sqrt{\frac{2}{3}}\M \tilde{H}_{,\phi\phi}}{\tilde{H}+\sqrt{\frac{2}{3}}\M \tilde{H}_{,\phi}}+\frac{\frac{\tilde{H}_{,\phi}}{\sqrt{6}\M}+\frac{4}{3}\tilde{H}_{,\phi\phi}+\sqrt{\frac{2}{3}}\M \tilde{H}_{,\phi\phi\phi}}{\frac{\tilde{H}}{\sqrt{6}\M}+\frac{4}{3}\tilde{H}_{,\phi}+\sqrt{\frac{2}{3}}\M \tilde{H}_{,\phi\phi}}}.
\end{split}
\ee
On using Eq.~(\ref{VGR}), one obtains the potential (\ref{VfRef}) from
\be{HfRef}
\tilde{H}(\phi)=h_a \re^{a\sqrt{\frac{3}{2}}\frac{\phi-\phi_0}{\M}}+h_b \re^{b\sqrt{\frac{3}{2}}\frac{\phi-\phi_0}{\M}},
\ee
if the constants $a$ and $b$ take certain values (see  the list of such values in Ref.~\cite{Vernov:2019ubo} and Table~\ref{table1}).
Therefore, the reconstruction can be analytically performed only in certain cases.
For example, for $a={}-2/3$  and $b=0$, we get
\begin{equation}\label{StarLambda}
    f(R)=\frac{\M^2}{2}\left(\frac{1}{24h_b^2}\,R^2-\frac{h_a}{h_b}R+\frac{8}{3}h_a^2\right)\,.
\end{equation}
Let $\phi_0$ be an attractor in the EF, then $\ep{1}=0$ if $\tilde{H}_{,\phi}(\phi_0)=0$ and in order for $\ep{2}\rightarrow {\rm const}$ at $\phi_0$ one needs $\tilde{H}_{,\phi\phi}(\phi_0)\neq 0$ and $\tilde{H}(\phi_0)>0$. On imposing such conditions on the expression (\ref{HfRef}), one obtains
\be{condfR}
\tilde{H}_{,\phi}(\phi_0)=\sqrt{\frac{3}{2}}\pa{\frac{a\,h_a}{\M}+\frac{b\,h_b}{\M}}=0\Rightarrow h_b={}-\frac{a}{b}h_a
\ee
and
\be{cond2fR}
\tilde{H}(\phi_0)= \frac{\pa{b-a}h_a}{b},\; \tilde{H}_{,\phi\phi}(\phi_0)=\frac{3a\pa{a-b}\,h_a}{2\M^2}.
\ee
Since we are interested in solutions where $\phi_0$ is an attractor, then close to $\phi_0$
\be{attcond}
\tilde G(\phi)\equiv\frac{\rd \phi}{\rd \tilde t} \sim -\tilde \alpha(\phi-\phi_0)
\ee
with $\tilde \alpha>0$. From  (\ref{acceq2}) one then has
\be{attcond2}
\tilde G(\phi)\sim-2\M^2 \tilde{H}_{,\phi\phi}(\phi_0)\pa{\phi-\phi_0}=3 a b \tilde{H}(\phi_0) (\phi-\phi_0)
\ee
and $\tilde\alpha={}-3 a b \tilde{H}(\phi_0)$ is positive if $a$ and $b$ have opposite signs [and we only consider solutions with a de Sitter attractor, $\tilde{H}(\phi_0)>0$].
Summarizing, the above requirement restricts the solutions to those having $h_a>0$, $(b-a)/b>0$ and $a\, b<0$. Such solutions are listed in  Table~\ref{table1} with  $W_{a}\equiv h_{a} \exp\pa{-a\sqrt{3/2}\,\phi_0/\M}$.  Values of $\epsilon_2$ in Table~\ref{table1} are calculated in the limit $\epsilon_1\rightarrow
0$.

\begin{table}[t!]
  \begin{center}
    \caption{Reconstructed $f(R)$ models}
    \label{table1}
    \begin{tabular}{|c|c|c|c|c|}
    \hline
      $a$ & $b$ & $\ep{2}^{(1)}$ & $\ep{2}^{(2)}$ & $2f(R)/\M^2$\\
      \hline
      \hline
      $-\frac{2}{3}$& $1$ &  $-2$ & $-1$ & $20 W_a^2  \pa{\frac{3R}{100} W_a^2}^{5/3}-\frac{10}{3}W_a^2$ \\
      \hline
      $-1$& $\frac{1}{3}$ &  $-2$ & $-1$ & $96W_a^2 \pa{\frac{R}{144 W_a^2}-\frac{1}{3}}^{3/2}$ \\
      \hline
      $-\frac{1}{3}$& $1$ &  $-2$ & $-1$ & $\frac{32}{3} W_a^2 \pa{\frac{R}{16 W_a^2}-\frac{1}{3}}^{3/2}$ \\
      \hline
      $-\frac{7}{3}$& $1$ &  $-7$ & $-$ & $160 W_a^2 \pa{\frac{3R}{400 W_a^2}}^{5/6}-\frac{280}{3} W_a^2$ \\
      \hline
      $-\frac{5}{3}$& $1$ &  $-5$ & $-$ & $\frac{128}{3} W_a^2 \pa{\frac{R}{32 W_a^2}-\frac{5}{3}}^{3/4}$ \\
      \hline
    \end{tabular}
  \end{center}
\end{table}

\section{de Sitter solutions in $f(R)$ theories}

\subsection{Stability}
Given the limitations for applying the reconstruction method to $f(R)$ theories, we henceforth adopt a less general approach compared with what we illustrated so far. Such an approach consists in studying some of the specific models discussed in the literature and verifying for each of them the existence of a de Sitter attractor and the existence of CR-like solutions leading to an enhancement of the primordial spectra. Indeed, a plethora of $f(R)$ theories has been proposed in order to explain inflation or the present cosmic acceleration and the existence of a de Sitter solution is crucial in order to possibly have the amplification.
In particular, let the attractor be $R=R_{0}>0$. The existence of such an attractor is consistent with Eq. (\ref{frfR}), which can be recast in the form
\be{frfR2}
1=\pa{2-\ep{1}}\frac{n_f-1}{n_f}-\frac{\dot R}{H R}n_F\,,\quad {\rm where}\quad n_g\equiv \frac{R}{g}\frac{\rd g}{\rd R},
\ee
hence,
\be{exdS}
n_f(R_0)=2.
\ee
From Eq.~(\ref{frfR2})
and the expression
\begin{equation*}
\dot{R}={}-6\epsilon_1H^3\left[\epsilon_2+2\,(2-\epsilon_1)\right],
\end{equation*}
it follows that
\begin{equation}
\label{epsilon2fR}  \epsilon_2=\left(2-\epsilon_1\right)  \left(\frac{n_f \left(\epsilon_1-1 \right) +2-\epsilon_1}{\epsilon_1\, n_F\,n_f}-2 \right).
\end{equation}
The stability of the solution with respect to homogeneous perturbations can be studied~\cite{Faraoni:2007yn} starting from the KG-like equation
\be{kgfR}
\ddot F+3 H \dot F+\frac{2f-RF}{3}=0
\ee
which can be easily obtained by differentiating (\ref{frfR2}). In Eq.~(\ref{kgfR}), analogously to the minimally coupled case, $F$ plays the role of the field and the last term is the potential gradient $V_{,F}$. The stability of a de Sitter solution (which indeed can be found on solving for $V_{,F}=0$) is equivalent to the condition $V_{,FF}>0$. This last condition becomes
\be{stabdSVFF}
V_{,FF}=\frac{\rd R}{\rd F}\frac{\rd V_{,F}}{\rd R}=R\frac{1-n_F}{3n_F}>0\Rightarrow 0<\left.n_F\right|_{R=R_0}<1\,,
\ee
provided $R_0>0$. We note that $\ep{1}=2-R/(6H^2)$, hence, $R>0$ during inflation.

The stability of de Sitter solutions can be also studied by using the equivalent scalar-tensor action (\ref{actfRJF}).
It has been shown that a de Sitter solution corresponds to the extrema of the following potential
\begin{equation}
            W\equiv\frac{RF-f}{F^2},
\end{equation}
If we only consider models that can be transformed to the EF model with a standard scalar field, hence assuming $F(R)>0$, de Sitter solutions are the solutions of
\begin{equation}\label{Vfprime}
    W_{,R}=\frac{(2f-RF)F_{,R}}{F^3}=0
\end{equation}
and the stable de Sitter solutions correspond to minima of $W$ (whereas unstable ones correspond to maxima~\cite{Vernov:2021hxo}).
It can be easily shown that this stability condition is equivalent to Eq.~(\ref{stabdSVFF}).


In order to calculate the second SR parameter $\ep{2}$ we consider expression (\ref{epsilon2fR}) in the limit $\ep{1}\rightarrow 0$ and $n_f\rightarrow 2$ and we get the equation
\be{eqE2}
\epsilon_2^2+3\epsilon_2-4\left(1-n_F^{-1}\right)=0\,,
\ee
that has the following two solutions for $\ep{2}$:
\be{ep2fR}
\ep{2}^{(1,2)}={}-\frac{1}{2}\pa{3\pm\sqrt{25-\frac{16}{n_F}\,}\,}\,.
\ee
 We observe that, provided the stability condition (\ref{stabdSVFF}) is satisfied, the limit for $\ep{2}$ exists only when $n_F\ge16/25$. In particular, one has that $-3<\ep{2}^{(1)}<-3/2$ and $-3/2<\ep{2}^{(2)}<0$.\\
If $0<n_F<16/25$, the solutions are stable, $\ep{1}\rightarrow 0$ and $\ep{2}$ has an oscillatory behavior with a constant amplitude. Even if this behavior may have interesting consequences on the shape of the inflationary spectra, we shall not discuss it further in the present article.

If the stability condition is not satisfied, one still has a solution with decreasing $\ep{1}$ corresponding to $\ep{2}^{(1)}$. In particular, for $n_F>1$ one finds $-3<\ep{2}^{(1)}<-4$ and for $n_F<0$ then $-\infty<\ep{2}^{(1)}<-4$.

Let us finally note that, for large times and close to the attractor one has
\be{nsm1f_R}
\Phi^{(1,2)}=3+2\ep{2}^{(1,2)}=\mp\sqrt{25-\frac{16}{n_F}}
\ee
and
\be{nsm1fR}
n_s^{(1,2)}-1=-2\ep{2}^{(1,2)}=3\pm\sqrt{25-\frac{16}{n_F}}.
\ee
Therefore, the Mukhanov--Sasaki equation has an increasing solution ($\Phi<0$) if $\ep{2}<-3/2$. If $-3/2<\ep{2}<0$ increasing solutions do not exist and the spectral index is blue-tilted.

\subsection{Scale invariant model}
The easiest model for amplification is $f(R)=\alpha R^2$ with $n_f=2$ and $n_F=1$ exactly. This implies that $R_0$ can take all positive values. For such a case, one finds
\be{ep2_fR1}
\ep{2}=-3
\ee
and $\Phi=-3<0$. The presence of an increasing solution for the curvature perturbations determines the amplification of the spectrum. The stability condition (\ref{stabdSVFF}) is not satisfied analogously to what occurs for the USR inflation and indeed the second solution to $\ep{2}$ corresponds to $\ep{2}=0$.
Let us note that the corresponding model in the EF has a constant potential.

\subsection{Second-order polynomial in $R$}
Let us now consider the model $f(R)=\alpha_0 M^4 +\alpha_1 M^2 R+\alpha_2 R^2$. For some values of the constants $\alpha_i$, we get model (\ref{StarLambda}) with an exact solution.
The de Sitter solution exists if $\alpha_0\alpha_1<0$ and is
\be{solnf2}
R_0=-2M^2\frac{\alpha_0}{\alpha_1}.
\ee
Correspondingly, one finds
\be{nF2}
n_F=\frac{1}{1-\frac{\alpha_1^2}{4\alpha_0\alpha_2}}
\ee
and the stability condition is satisfied for $0<\alpha_1^2/(4\alpha_0\alpha_2)<1$.
\subsection{$n$-order trinomial}
Let us now consider the following, more general form of $f$:
\be{trinf}
f(R)=M^2 R+\alpha \,R^2 +\beta \,\frac{R^n}{M^{2n}}.
\ee
The de Sitter solution exists if $\beta(n-2)>0$ and is
\be{solnf3}
R_0=M^2\paq{\frac{1}{\beta(n-2)}}^{1/(n-1)}.
\ee
Correspondingly, one obtains
\be{nF3}
n_F=
\frac{1+\frac{n}{2}\nu}{1+\nu}\,,\quad{\rm where}\quad \nu=\frac{M^2(n-1)}{R_0\alpha (n-2)}\,.
\ee
Stability of the de Sitter solutions in models described by (\ref{trinf}) have been considered in~\cite{Pham:2024fub}. Using condition (\ref{stabdSVFF}), we distinguish three different cases. If $0<n<2$, the stability condition requires $\nu>0$ or $\nu<-2/n$. If $n>2$, then the de Sitter solution is stable at $-2/n<\nu<0$. Finally, for $n<0$, one needs $0<\nu<-2/n$ for stability.

When $\alpha=0$,
the general expression (\ref{trinf})  becomes a binomial with different powers of~$R$.
For such a case, 
one finds
\be{nF1}
n_F=\frac{n}{2}
\ee
and stable solutions therefore exist if $0<n<2$ that corresponds to $\beta<0$. When $n>2$ and $n<0$ the solution (\ref{solnf3}) exists but is not stable.
We also remind the reader that the condition $F(R_0)>0$ should be satisfied.
In order to have a definite limit for $\ep{2}$ one also needs $n\ge 32/25$ and the stable solutions lead to
\be{phinsm11}
\Phi=\mp\sqrt{25-\frac{32}{n}}\quad{\rm and}\quad n_s-1=3\pm\sqrt{25-\frac{32}{n}}.
\ee
Let us note that for $n>2$ and $n<0$ one solution which approaches the de Sitter limit still exists and has
\be{ep2fR1}
\ep{2}=-\frac{1}{2}\pa{3+\sqrt{25-\frac{32}{n}}}.
\ee

\section{Conclusions}
Nowadays, primordial black holes are very attractive candidates to explain dark matter~\cite{Carr:2020xqk,Carr:2023tpt,Espinosa:2017sgp} and the origin of the large masses of the BH binaries was observed by the LIGO/Virgo collaborations a few years ago. Furthermore, PBHs may explain the spectrum of the stochastic GW background which has recently been observed by NanoGRAV and other experiments and give further insights on the physics of inflation.

Indeed, large primordial curvature perturbations can be generated during inflation and collapse after inflation ends forming PBHs. Several mechanisms to amplify the inflationary perturbations are currently studied. An inflationary period with a phase of USR or CR at energy scales smaller than those probed by CMB has been considered in diverse inflationary models. During such a phase the homogeneous inflaton-gravity system rapidly evolves towards a de Sitter attractor wherein the Mukhanov--Sasaki equation governing the evolution of the curvature perturbation has a growing solution or may generate a blue-tilted spectrum. In this paper, we discussed the general features of some of the inflationary models which employ this mechanism to generate the perturbations enhancement. In particular, whenever possible, we applied the superpotential method to reconstruct the inflaton potential generating an USR or a CR evolution. These inflaton potential features are found for a minimally coupled inflaton field and in some modified gravity models involving a nonminimal coupling and $f(R)$ theories. This article is the generalization of a previous work~\cite{Chataignier:2023ago} wherein a different technique was employed for the reconstruction. We found that the superpotential method is a far more powerful method in this context, predicting the shape of the inflaton potential close to the attractor for a large class of models. The stability of the evolution in such models is also studied. Moreover, reconstruction could also be applied to $f(R)$ theories, but has a restricted range of applicability in such cases. Therefore, in such a context, besides illustrating how reconstruction could work, a set of $f(R)$ inflationary models was also considered and discussed as a possible source of an amplification of the primordial spectra.


\begin{thebibliography}{67}
\expandafter\ifx\csname natexlab\endcsname\relax\def\natexlab#1{#1}\fi
\expandafter\ifx\csname bibnamefont\endcsname\relax
  \def\bibnamefont#1{#1}\fi
\expandafter\ifx\csname bibfnamefont\endcsname\relax
  \def\bibfnamefont#1{#1}\fi
\expandafter\ifx\csname citenamefont\endcsname\relax
  \def\citenamefont#1{#1}\fi
\expandafter\ifx\csname url\endcsname\relax
  \def\url#1{\texttt{#1}}\fi
\expandafter\ifx\csname urlprefix\endcsname\relax\def\urlprefix{URL }\fi
\providecommand{\bibinfo}[2]{#2}
\providecommand{\eprint}[2][]{\url{#2}}

\bibitem[{\citenamefont{Zel'dovich and Novikov}(1967)}]{Zeldovich:1967lct}
\bibinfo{author}{\bibfnamefont{Y.~B.} \bibnamefont{Zel'dovich}}
  \bibnamefont{and} \bibinfo{author}{\bibfnamefont{I.~D.}
  \bibnamefont{Novikov}}, \bibinfo{journal}{Sov. Astron.}
  \textbf{\bibinfo{volume}{10}}, \bibinfo{pages}{602} (\bibinfo{year}{1967}).

\bibitem[{\citenamefont{Hawking}(1971)}]{Hawking:1971ei}
\bibinfo{author}{\bibfnamefont{S.}~\bibnamefont{Hawking}},
  \bibinfo{journal}{Mon. Not. Roy. Astron. Soc.}
  \textbf{\bibinfo{volume}{152}}, \bibinfo{pages}{75} (\bibinfo{year}{1971}).

\bibitem[{\citenamefont{Abbott et~al.}(2016)}]{LIGOScientific:2016aoc}
\bibinfo{author}{\bibfnamefont{B.~P.} \bibnamefont{Abbott}}
  \bibnamefont{et~al.} (\bibinfo{collaboration}{LIGO Scientific, Virgo}),
  \bibinfo{journal}{Phys. Rev. Lett.} \textbf{\bibinfo{volume}{116}},
  \bibinfo{pages}{061102} (\bibinfo{year}{2016}), \eprint{1602.03837}.

\bibitem[{\citenamefont{Dolgov}(2018)}]{Dolgov:2017aec}
\bibinfo{author}{\bibfnamefont{A.~D.} \bibnamefont{Dolgov}},
  \bibinfo{journal}{Usp. Fiz. Nauk} \textbf{\bibinfo{volume}{188}},
  \bibinfo{pages}{121} (\bibinfo{year}{2018}), \eprint{1705.06859}.

\bibitem[{\citenamefont{Carr and Kuhnel}(2020)}]{Carr:2020xqk}
\bibinfo{author}{\bibfnamefont{B.}~\bibnamefont{Carr}} \bibnamefont{and}
  \bibinfo{author}{\bibfnamefont{F.}~\bibnamefont{Kuhnel}},
  \bibinfo{journal}{Ann. Rev. Nucl. Part. Sci.} \textbf{\bibinfo{volume}{70}},
  \bibinfo{pages}{355} (\bibinfo{year}{2020}), \eprint{2006.02838}.

\bibitem[{\citenamefont{\"Ozsoy and Tasinato}(2023)}]{Ozsoy:2023ryl}
\bibinfo{author}{\bibfnamefont{O.}~\bibnamefont{\"Ozsoy}} \bibnamefont{and}
  \bibinfo{author}{\bibfnamefont{G.}~\bibnamefont{Tasinato}},
  \bibinfo{journal}{Universe} \textbf{\bibinfo{volume}{9}},
  \bibinfo{pages}{203} (\bibinfo{year}{2023}), \eprint{2301.03600}.

\bibitem[{\citenamefont{Carr et~al.}(2024)\citenamefont{Carr, Clesse,
  Garcia-Bellido, Hawkins, and Kuhnel}}]{Carr:2023tpt}
\bibinfo{author}{\bibfnamefont{B.}~\bibnamefont{Carr}},
  \bibinfo{author}{\bibfnamefont{S.}~\bibnamefont{Clesse}},
  \bibinfo{author}{\bibfnamefont{J.}~\bibnamefont{Garcia-Bellido}},
  \bibinfo{author}{\bibfnamefont{M.}~\bibnamefont{Hawkins}}, \bibnamefont{and}
  \bibinfo{author}{\bibfnamefont{F.}~\bibnamefont{Kuhnel}},
  \bibinfo{journal}{Phys. Rept.} \textbf{\bibinfo{volume}{1054}},
  \bibinfo{pages}{1} (\bibinfo{year}{2024}), \eprint{2306.03903}.

\bibitem[{\citenamefont{Espinosa et~al.}(2018)\citenamefont{Espinosa, Racco,
  and Riotto}}]{Espinosa:2017sgp}
\bibinfo{author}{\bibfnamefont{J.~R.} \bibnamefont{Espinosa}},
  \bibinfo{author}{\bibfnamefont{D.}~\bibnamefont{Racco}}, \bibnamefont{and}
  \bibinfo{author}{\bibfnamefont{A.}~\bibnamefont{Riotto}},
  \bibinfo{journal}{Phys. Rev. Lett.} \textbf{\bibinfo{volume}{120}},
  \bibinfo{pages}{121301} (\bibinfo{year}{2018}), \eprint{1710.11196}.

\bibitem[{\citenamefont{Dolgov and Silk}(1993)}]{Dolgov:1992pu}
\bibinfo{author}{\bibfnamefont{A.}~\bibnamefont{Dolgov}} \bibnamefont{and}
  \bibinfo{author}{\bibfnamefont{J.}~\bibnamefont{Silk}},
  \bibinfo{journal}{Phys. Rev. D} \textbf{\bibinfo{volume}{47}},
  \bibinfo{pages}{4244} (\bibinfo{year}{1993}).

\bibitem[{\citenamefont{Ivanov et~al.}(1994)\citenamefont{Ivanov, Naselsky, and
  Novikov}}]{Ivanov:1994pa}
\bibinfo{author}{\bibfnamefont{P.}~\bibnamefont{Ivanov}},
  \bibinfo{author}{\bibfnamefont{P.}~\bibnamefont{Naselsky}}, \bibnamefont{and}
  \bibinfo{author}{\bibfnamefont{I.}~\bibnamefont{Novikov}},
  \bibinfo{journal}{Phys. Rev. D} \textbf{\bibinfo{volume}{50}},
  \bibinfo{pages}{7173} (\bibinfo{year}{1994}).

\bibitem[{\citenamefont{Garcia-Bellido
  et~al.}(1996)\citenamefont{Garcia-Bellido, Linde, and
  Wands}}]{Garcia-Bellido:1996mdl}
\bibinfo{author}{\bibfnamefont{J.}~\bibnamefont{Garcia-Bellido}},
  \bibinfo{author}{\bibfnamefont{A.~D.} \bibnamefont{Linde}}, \bibnamefont{and}
  \bibinfo{author}{\bibfnamefont{D.}~\bibnamefont{Wands}},
  \bibinfo{journal}{Phys. Rev. D} \textbf{\bibinfo{volume}{54}},
  \bibinfo{pages}{6040} (\bibinfo{year}{1996}), \eprint{astro-ph/9605094}.

\bibitem[{\citenamefont{Karam et~al.}(2023)\citenamefont{Karam, Koivunen,
  Tomberg, Vaskonen, and Veerm\"ae}}]{Karam:2022nym}
\bibinfo{author}{\bibfnamefont{A.}~\bibnamefont{Karam}},
  \bibinfo{author}{\bibfnamefont{N.}~\bibnamefont{Koivunen}},
  \bibinfo{author}{\bibfnamefont{E.}~\bibnamefont{Tomberg}},
  \bibinfo{author}{\bibfnamefont{V.}~\bibnamefont{Vaskonen}}, \bibnamefont{and}
  \bibinfo{author}{\bibfnamefont{H.}~\bibnamefont{Veerm\"ae}},
  \bibinfo{journal}{JCAP} \textbf{\bibinfo{volume}{03}}, \bibinfo{pages}{013}
  (\bibinfo{year}{2023}), \eprint{2205.13540}.

\bibitem[{\citenamefont{Kristiano and Yokoyama}(2024)}]{Kristiano:2022maq}
\bibinfo{author}{\bibfnamefont{J.}~\bibnamefont{Kristiano}} \bibnamefont{and}
  \bibinfo{author}{\bibfnamefont{J.}~\bibnamefont{Yokoyama}},
  \bibinfo{journal}{Phys. Rev. Lett.} \textbf{\bibinfo{volume}{132}},
  \bibinfo{pages}{221003} (\bibinfo{year}{2024}), \eprint{2211.03395}.

\bibitem[{\citenamefont{Ketov}(2021)}]{Ketov:2021fww}
\bibinfo{author}{\bibfnamefont{S.~V.} \bibnamefont{Ketov}},
  \bibinfo{journal}{Universe} \textbf{\bibinfo{volume}{7}},
  \bibinfo{pages}{115} (\bibinfo{year}{2021}).

\bibitem[{\citenamefont{Muslimov}(1990)}]{Muslimov:1990be}
\bibinfo{author}{\bibfnamefont{A.~G.} \bibnamefont{Muslimov}},
  \bibinfo{journal}{Class. Quant. Grav.} \textbf{\bibinfo{volume}{7}},
  \bibinfo{pages}{231} (\bibinfo{year}{1990}).

\bibitem[{\citenamefont{Salopek and Bond}(1990)}]{Salopek:1990jq}
\bibinfo{author}{\bibfnamefont{D.~S.} \bibnamefont{Salopek}} \bibnamefont{and}
  \bibinfo{author}{\bibfnamefont{J.~R.} \bibnamefont{Bond}},
  \bibinfo{journal}{Phys. Rev. D} \textbf{\bibinfo{volume}{42}},
  \bibinfo{pages}{3936} (\bibinfo{year}{1990}).

\bibitem[{\citenamefont{Aref'eva et~al.}(2006)\citenamefont{Aref'eva, Koshelev,
  and Vernov}}]{Arefeva:2004odl}
\bibinfo{author}{\bibfnamefont{I.~Y.} \bibnamefont{Aref'eva}},
  \bibinfo{author}{\bibfnamefont{A.~S.} \bibnamefont{Koshelev}},
  \bibnamefont{and} \bibinfo{author}{\bibfnamefont{S.~Y.}
  \bibnamefont{Vernov}}, \bibinfo{journal}{Theor. Math. Phys.}
  \textbf{\bibinfo{volume}{148}}, \bibinfo{pages}{895} (\bibinfo{year}{2006}),
  \eprint{astro-ph/0412619}.

\bibitem[{\citenamefont{Townsend}(2008)}]{Townsend:2007aw}
\bibinfo{author}{\bibfnamefont{P.~K.} \bibnamefont{Townsend}},
  \bibinfo{journal}{Class. Quant. Grav.} \textbf{\bibinfo{volume}{25}},
  \bibinfo{pages}{045017} (\bibinfo{year}{2008}), \eprint{0710.5178}.

\bibitem[{\citenamefont{Kamenshchik et~al.}(2013)\citenamefont{Kamenshchik,
  Tronconi, Venturi, and Vernov}}]{Kamenshchik:2012pw}
\bibinfo{author}{\bibfnamefont{A.~Y.} \bibnamefont{Kamenshchik}},
  \bibinfo{author}{\bibfnamefont{A.}~\bibnamefont{Tronconi}},
  \bibinfo{author}{\bibfnamefont{G.}~\bibnamefont{Venturi}}, \bibnamefont{and}
  \bibinfo{author}{\bibfnamefont{S.~Y.} \bibnamefont{Vernov}},
  \bibinfo{journal}{Phys. Rev. D} \textbf{\bibinfo{volume}{87}},
  \bibinfo{pages}{063503} (\bibinfo{year}{2013}), \eprint{1211.6272}.

\bibitem[{\citenamefont{Vernov et~al.}(2020)\citenamefont{Vernov, Ivanov, and
  Pozdeeva}}]{Vernov:2019ubo}
\bibinfo{author}{\bibfnamefont{S.~Y.} \bibnamefont{Vernov}},
  \bibinfo{author}{\bibfnamefont{V.~R.} \bibnamefont{Ivanov}},
  \bibnamefont{and} \bibinfo{author}{\bibfnamefont{E.~O.}
  \bibnamefont{Pozdeeva}}, \bibinfo{journal}{Phys. Part. Nucl.}
  \textbf{\bibinfo{volume}{51}}, \bibinfo{pages}{744} (\bibinfo{year}{2020}),
  \eprint{1912.07049}.

\bibitem[{\citenamefont{Motohashi et~al.}(2015)\citenamefont{Motohashi,
  Starobinsky, and Yokoyama}}]{Motohashi:2014ppa}
\bibinfo{author}{\bibfnamefont{H.}~\bibnamefont{Motohashi}},
  \bibinfo{author}{\bibfnamefont{A.~A.} \bibnamefont{Starobinsky}},
  \bibnamefont{and} \bibinfo{author}{\bibfnamefont{J.}~\bibnamefont{Yokoyama}},
  \bibinfo{journal}{JCAP} \textbf{\bibinfo{volume}{09}}, \bibinfo{pages}{018}
  (\bibinfo{year}{2015}), \eprint{1411.5021}.

\bibitem[{\citenamefont{Nojiri et~al.}(2017)\citenamefont{Nojiri, Odintsov, and
  Oikonomou}}]{Nojiri:2017qvx}
\bibinfo{author}{\bibfnamefont{S.}~\bibnamefont{Nojiri}},
  \bibinfo{author}{\bibfnamefont{S.~D.} \bibnamefont{Odintsov}},
  \bibnamefont{and} \bibinfo{author}{\bibfnamefont{V.~K.}
  \bibnamefont{Oikonomou}}, \bibinfo{journal}{Class. Quant. Grav.}
  \textbf{\bibinfo{volume}{34}}, \bibinfo{pages}{245012}
  (\bibinfo{year}{2017}), \eprint{1704.05945}.

\bibitem[{\citenamefont{Motohashi and Starobinsky}(2017)}]{Motohashi:2017vdc}
\bibinfo{author}{\bibfnamefont{H.}~\bibnamefont{Motohashi}} \bibnamefont{and}
  \bibinfo{author}{\bibfnamefont{A.~A.} \bibnamefont{Starobinsky}},
  \bibinfo{journal}{Eur. Phys. J. C} \textbf{\bibinfo{volume}{77}},
  \bibinfo{pages}{538} (\bibinfo{year}{2017}), \eprint{1704.08188}.

\bibitem[{\citenamefont{Yi and Gong}(2018)}]{Yi:2017mxs}
\bibinfo{author}{\bibfnamefont{Z.}~\bibnamefont{Yi}} \bibnamefont{and}
  \bibinfo{author}{\bibfnamefont{Y.}~\bibnamefont{Gong}},
  \bibinfo{journal}{JCAP} \textbf{\bibinfo{volume}{03}}, \bibinfo{pages}{052}
  (\bibinfo{year}{2018}), \eprint{1712.07478}.

\bibitem[{\citenamefont{Chataignier et~al.}(2023)\citenamefont{Chataignier,
  Kamenshchik, Tronconi, and Venturi}}]{Chataignier:2023ago}
\bibinfo{author}{\bibfnamefont{L.}~\bibnamefont{Chataignier}},
  \bibinfo{author}{\bibfnamefont{A.~Y.} \bibnamefont{Kamenshchik}},
  \bibinfo{author}{\bibfnamefont{A.}~\bibnamefont{Tronconi}}, \bibnamefont{and}
  \bibinfo{author}{\bibfnamefont{G.}~\bibnamefont{Venturi}},
  \bibinfo{journal}{Phys. Rev. D} \textbf{\bibinfo{volume}{107}},
  \bibinfo{pages}{083506} (\bibinfo{year}{2023}), \eprint{2301.04477}.

\bibitem[{\citenamefont{Liddle et~al.}(1994)\citenamefont{Liddle, Parsons, and
  Barrow}}]{Liddle:1994dx}
\bibinfo{author}{\bibfnamefont{A.~R.} \bibnamefont{Liddle}},
  \bibinfo{author}{\bibfnamefont{P.}~\bibnamefont{Parsons}}, \bibnamefont{and}
  \bibinfo{author}{\bibfnamefont{J.~D.} \bibnamefont{Barrow}},
  \bibinfo{journal}{Phys. Rev. D} \textbf{\bibinfo{volume}{50}},
  \bibinfo{pages}{7222} (\bibinfo{year}{1994}), \eprint{astro-ph/9408015}.

\bibitem[{\citenamefont{Kinney}(1997)}]{Kinney:1997ne}
\bibinfo{author}{\bibfnamefont{W.~H.} \bibnamefont{Kinney}},
  \bibinfo{journal}{Phys. Rev. D} \textbf{\bibinfo{volume}{56}},
  \bibinfo{pages}{2002} (\bibinfo{year}{1997}), \eprint{hep-ph/9702427}.

\bibitem[{\citenamefont{Starobinsky}(1980)}]{Starobinsky:1980te}
\bibinfo{author}{\bibfnamefont{A.~A.} \bibnamefont{Starobinsky}},
  \bibinfo{journal}{Phys. Lett. B} \textbf{\bibinfo{volume}{91}},
  \bibinfo{pages}{99} (\bibinfo{year}{1980}).

\bibitem[{\citenamefont{Maeda}(1988)}]{Maeda:1987xf}
\bibinfo{author}{\bibfnamefont{K.-i.} \bibnamefont{Maeda}},
  \bibinfo{journal}{Phys. Rev. D} \textbf{\bibinfo{volume}{37}},
  \bibinfo{pages}{858} (\bibinfo{year}{1988}).

\bibitem[{\citenamefont{Berkin and Maeda}(1990)}]{Berkin:1990nu}
\bibinfo{author}{\bibfnamefont{A.~L.} \bibnamefont{Berkin}} \bibnamefont{and}
  \bibinfo{author}{\bibfnamefont{K.-i.} \bibnamefont{Maeda}},
  \bibinfo{journal}{Phys. Lett. B} \textbf{\bibinfo{volume}{245}},
  \bibinfo{pages}{348} (\bibinfo{year}{1990}).

\bibitem[{\citenamefont{Huang}(2014)}]{Huang:2013hsb}
\bibinfo{author}{\bibfnamefont{Q.-G.} \bibnamefont{Huang}},
  \bibinfo{journal}{JCAP} \textbf{\bibinfo{volume}{02}}, \bibinfo{pages}{035}
  (\bibinfo{year}{2014}), \eprint{1309.3514}.

\bibitem[{\citenamefont{Motohashi}(2015)}]{Motohashi:2014tra}
\bibinfo{author}{\bibfnamefont{H.}~\bibnamefont{Motohashi}},
  \bibinfo{journal}{Phys. Rev. D} \textbf{\bibinfo{volume}{91}},
  \bibinfo{pages}{064016} (\bibinfo{year}{2015}), \eprint{1411.2972}.

\bibitem[{\citenamefont{Miranda et~al.}(2017)\citenamefont{Miranda, Fabris, and
  Piattella}}]{Miranda:2017juz}
\bibinfo{author}{\bibfnamefont{T.}~\bibnamefont{Miranda}},
  \bibinfo{author}{\bibfnamefont{J.~C.} \bibnamefont{Fabris}},
  \bibnamefont{and} \bibinfo{author}{\bibfnamefont{O.~F.}
  \bibnamefont{Piattella}}, \bibinfo{journal}{JCAP}
  \textbf{\bibinfo{volume}{09}}, \bibinfo{pages}{041} (\bibinfo{year}{2017}),
  \eprint{1707.06457}.

\bibitem[{\citenamefont{Ketov}(2020)}]{Ketov:2019toi}
\bibinfo{author}{\bibfnamefont{S.~V.} \bibnamefont{Ketov}},
  \bibinfo{journal}{J. Phys. A} \textbf{\bibinfo{volume}{53}},
  \bibinfo{pages}{084001} (\bibinfo{year}{2020}), \eprint{1911.01008}.

\bibitem[{\citenamefont{Ivanov et~al.}(2022)\citenamefont{Ivanov, Ketov,
  Pozdeeva, and Vernov}}]{Ivanov:2021chn}
\bibinfo{author}{\bibfnamefont{V.~R.} \bibnamefont{Ivanov}},
  \bibinfo{author}{\bibfnamefont{S.~V.} \bibnamefont{Ketov}},
  \bibinfo{author}{\bibfnamefont{E.~O.} \bibnamefont{Pozdeeva}},
  \bibnamefont{and} \bibinfo{author}{\bibfnamefont{S.~Y.}
  \bibnamefont{Vernov}}, \bibinfo{journal}{JCAP} \textbf{\bibinfo{volume}{03}},
  \bibinfo{pages}{058} (\bibinfo{year}{2022}), \eprint{2111.09058}.

\bibitem[{\citenamefont{Frolovsky et~al.}(2022)\citenamefont{Frolovsky, Ketov,
  and Saburov}}]{Frolovsky:2022ewg}
\bibinfo{author}{\bibfnamefont{D.}~\bibnamefont{Frolovsky}},
  \bibinfo{author}{\bibfnamefont{S.~V.} \bibnamefont{Ketov}}, \bibnamefont{and}
  \bibinfo{author}{\bibfnamefont{S.}~\bibnamefont{Saburov}},
  \bibinfo{journal}{Mod. Phys. Lett. A} \textbf{\bibinfo{volume}{37}},
  \bibinfo{pages}{2250135} (\bibinfo{year}{2022}), \eprint{2205.00603}.

\bibitem[{\citenamefont{Pozdeeva and Vernov}(2023)}]{Pozdeeva:2022lcj}
\bibinfo{author}{\bibfnamefont{E.~O.} \bibnamefont{Pozdeeva}} \bibnamefont{and}
  \bibinfo{author}{\bibfnamefont{S.~Y.} \bibnamefont{Vernov}},
  \bibinfo{journal}{Phys. Scripta} \textbf{\bibinfo{volume}{98}},
  \bibinfo{pages}{055001} (\bibinfo{year}{2023}), \eprint{2211.10988}.

\bibitem[{\citenamefont{Brinkmann et~al.}(2023)\citenamefont{Brinkmann, Cicoli,
  and Zito}}]{Brinkmann:2023eph}
\bibinfo{author}{\bibfnamefont{M.}~\bibnamefont{Brinkmann}},
  \bibinfo{author}{\bibfnamefont{M.}~\bibnamefont{Cicoli}}, \bibnamefont{and}
  \bibinfo{author}{\bibfnamefont{P.}~\bibnamefont{Zito}},
  \bibinfo{journal}{JHEP} \textbf{\bibinfo{volume}{09}}, \bibinfo{pages}{038}
  (\bibinfo{year}{2023}), \eprint{2305.05703}.

\bibitem[{\citenamefont{Saburov and Ketov}(2023)}]{Saburov:2023buy}
\bibinfo{author}{\bibfnamefont{S.}~\bibnamefont{Saburov}} \bibnamefont{and}
  \bibinfo{author}{\bibfnamefont{S.~V.} \bibnamefont{Ketov}},
  \bibinfo{journal}{Universe} \textbf{\bibinfo{volume}{9}},
  \bibinfo{pages}{323} (\bibinfo{year}{2023}), \eprint{2306.06597}.

\bibitem[{\citenamefont{Saburov and Ketov}(2024)}]{Saburov:2024und}
\bibinfo{author}{\bibfnamefont{S.}~\bibnamefont{Saburov}} \bibnamefont{and}
  \bibinfo{author}{\bibfnamefont{S.~V.} \bibnamefont{Ketov}}
  (\bibinfo{year}{2024}), \eprint{2402.02934}.

\bibitem[{\citenamefont{Spokoiny}(1984)}]{Spokoiny:1984bd}
\bibinfo{author}{\bibfnamefont{B.~L.} \bibnamefont{Spokoiny}},
  \bibinfo{journal}{Phys. Lett. B} \textbf{\bibinfo{volume}{147}},
  \bibinfo{pages}{39} (\bibinfo{year}{1984}).

\bibitem[{\citenamefont{Accetta et~al.}(1985)\citenamefont{Accetta, Zoller, and
  Turner}}]{Accetta:1985du}
\bibinfo{author}{\bibfnamefont{F.~S.} \bibnamefont{Accetta}},
  \bibinfo{author}{\bibfnamefont{D.~J.} \bibnamefont{Zoller}},
  \bibnamefont{and} \bibinfo{author}{\bibfnamefont{M.~S.}
  \bibnamefont{Turner}}, \bibinfo{journal}{Phys. Rev. D}
  \textbf{\bibinfo{volume}{31}}, \bibinfo{pages}{3046} (\bibinfo{year}{1985}).

\bibitem[{\citenamefont{Futamase and Maeda}(1989)}]{Futamase:1987ua}
\bibinfo{author}{\bibfnamefont{T.}~\bibnamefont{Futamase}} \bibnamefont{and}
  \bibinfo{author}{\bibfnamefont{K.-i.} \bibnamefont{Maeda}},
  \bibinfo{journal}{Phys. Rev. D} \textbf{\bibinfo{volume}{39}},
  \bibinfo{pages}{399} (\bibinfo{year}{1989}).

\bibitem[{\citenamefont{Barvinsky and Kamenshchik}(1994)}]{Barvinsky:1994hx}
\bibinfo{author}{\bibfnamefont{A.~O.} \bibnamefont{Barvinsky}}
  \bibnamefont{and} \bibinfo{author}{\bibfnamefont{A.~Y.}
  \bibnamefont{Kamenshchik}}, \bibinfo{journal}{Phys. Lett. B}
  \textbf{\bibinfo{volume}{332}}, \bibinfo{pages}{270} (\bibinfo{year}{1994}),
  \eprint{gr-qc/9404062}.

\bibitem[{\citenamefont{Cervantes-Cota and
  Dehnen}(1995)}]{Cervantes-Cota:1995ehs}
\bibinfo{author}{\bibfnamefont{J.~L.} \bibnamefont{Cervantes-Cota}}
  \bibnamefont{and} \bibinfo{author}{\bibfnamefont{H.}~\bibnamefont{Dehnen}},
  \bibinfo{journal}{Nucl. Phys. B} \textbf{\bibinfo{volume}{442}},
  \bibinfo{pages}{391} (\bibinfo{year}{1995}), \eprint{astro-ph/9505069}.

\bibitem[{\citenamefont{Bezrukov and Shaposhnikov}(2008)}]{Bezrukov:2007ep}
\bibinfo{author}{\bibfnamefont{F.~L.} \bibnamefont{Bezrukov}} \bibnamefont{and}
  \bibinfo{author}{\bibfnamefont{M.}~\bibnamefont{Shaposhnikov}},
  \bibinfo{journal}{Phys. Lett. B} \textbf{\bibinfo{volume}{659}},
  \bibinfo{pages}{703} (\bibinfo{year}{2008}), \eprint{0710.3755}.

\bibitem[{\citenamefont{Barvinsky et~al.}(2008)\citenamefont{Barvinsky,
  Kamenshchik, and Starobinsky}}]{Barvinsky:2008ia}
\bibinfo{author}{\bibfnamefont{A.~O.} \bibnamefont{Barvinsky}},
  \bibinfo{author}{\bibfnamefont{A.~Y.} \bibnamefont{Kamenshchik}},
  \bibnamefont{and} \bibinfo{author}{\bibfnamefont{A.~A.}
  \bibnamefont{Starobinsky}}, \bibinfo{journal}{JCAP}
  \textbf{\bibinfo{volume}{11}}, \bibinfo{pages}{021} (\bibinfo{year}{2008}),
  \eprint{0809.2104}.

\bibitem[{\citenamefont{Cerioni et~al.}(2009)\citenamefont{Cerioni, Finelli,
  Tronconi, and Venturi}}]{Cerioni:2009kn}
\bibinfo{author}{\bibfnamefont{A.}~\bibnamefont{Cerioni}},
  \bibinfo{author}{\bibfnamefont{F.}~\bibnamefont{Finelli}},
  \bibinfo{author}{\bibfnamefont{A.}~\bibnamefont{Tronconi}}, \bibnamefont{and}
  \bibinfo{author}{\bibfnamefont{G.}~\bibnamefont{Venturi}},
  \bibinfo{journal}{Phys. Lett. B} \textbf{\bibinfo{volume}{681}},
  \bibinfo{pages}{383} (\bibinfo{year}{2009}), \eprint{0906.1902}.

\bibitem[{\citenamefont{Ferrara et~al.}(2010)\citenamefont{Ferrara, Kallosh,
  Linde, Marrani, and Van~Proeyen}}]{Ferrara:2010yw}
\bibinfo{author}{\bibfnamefont{S.}~\bibnamefont{Ferrara}},
  \bibinfo{author}{\bibfnamefont{R.}~\bibnamefont{Kallosh}},
  \bibinfo{author}{\bibfnamefont{A.}~\bibnamefont{Linde}},
  \bibinfo{author}{\bibfnamefont{A.}~\bibnamefont{Marrani}}, \bibnamefont{and}
  \bibinfo{author}{\bibfnamefont{A.}~\bibnamefont{Van~Proeyen}},
  \bibinfo{journal}{Phys. Rev. D} \textbf{\bibinfo{volume}{82}},
  \bibinfo{pages}{045003} (\bibinfo{year}{2010}), \eprint{1004.0712}.

\bibitem[{\citenamefont{Elizalde et~al.}(2014)\citenamefont{Elizalde, Odintsov,
  Pozdeeva, and Vernov}}]{Elizalde:2014xva}
\bibinfo{author}{\bibfnamefont{E.}~\bibnamefont{Elizalde}},
  \bibinfo{author}{\bibfnamefont{S.~D.} \bibnamefont{Odintsov}},
  \bibinfo{author}{\bibfnamefont{E.~O.} \bibnamefont{Pozdeeva}},
  \bibnamefont{and} \bibinfo{author}{\bibfnamefont{S.~Y.}
  \bibnamefont{Vernov}}, \bibinfo{journal}{Phys. Rev. D}
  \textbf{\bibinfo{volume}{90}}, \bibinfo{pages}{084001}
  (\bibinfo{year}{2014}), \eprint{1408.1285}.

\bibitem[{\citenamefont{Elizalde et~al.}(2016)\citenamefont{Elizalde, Odintsov,
  Pozdeeva, and Vernov}}]{Elizalde:2015nya}
\bibinfo{author}{\bibfnamefont{E.}~\bibnamefont{Elizalde}},
  \bibinfo{author}{\bibfnamefont{S.~D.} \bibnamefont{Odintsov}},
  \bibinfo{author}{\bibfnamefont{E.~O.} \bibnamefont{Pozdeeva}},
  \bibnamefont{and} \bibinfo{author}{\bibfnamefont{S.~Y.}
  \bibnamefont{Vernov}}, \bibinfo{journal}{JCAP} \textbf{\bibinfo{volume}{02}},
  \bibinfo{pages}{025} (\bibinfo{year}{2016}), \eprint{1509.08817}.

\bibitem[{\citenamefont{Binetruy et~al.}(2015)\citenamefont{Binetruy, Kiritsis,
  Mabillard, Pieroni, and Rosset}}]{Binetruy:2014zya}
\bibinfo{author}{\bibfnamefont{P.}~\bibnamefont{Binetruy}},
  \bibinfo{author}{\bibfnamefont{E.}~\bibnamefont{Kiritsis}},
  \bibinfo{author}{\bibfnamefont{J.}~\bibnamefont{Mabillard}},
  \bibinfo{author}{\bibfnamefont{M.}~\bibnamefont{Pieroni}}, \bibnamefont{and}
  \bibinfo{author}{\bibfnamefont{C.}~\bibnamefont{Rosset}},
  \bibinfo{journal}{JCAP} \textbf{\bibinfo{volume}{04}}, \bibinfo{pages}{033}
  (\bibinfo{year}{2015}), \eprint{1407.0820}.

\bibitem[{\citenamefont{Carr}(1975)}]{Carr:1975qj}
\bibinfo{author}{\bibfnamefont{B.~J.} \bibnamefont{Carr}},
  \bibinfo{journal}{Astrophys. J.} \textbf{\bibinfo{volume}{201}},
  \bibinfo{pages}{1} (\bibinfo{year}{1975}).

\bibitem[{\citenamefont{Chen et~al.}(2013)\citenamefont{Chen, Firouzjahi,
  Namjoo, and Sasaki}}]{Chen:2013kta}
\bibinfo{author}{\bibfnamefont{X.}~\bibnamefont{Chen}},
  \bibinfo{author}{\bibfnamefont{H.}~\bibnamefont{Firouzjahi}},
  \bibinfo{author}{\bibfnamefont{M.~H.} \bibnamefont{Namjoo}},
  \bibnamefont{and} \bibinfo{author}{\bibfnamefont{M.}~\bibnamefont{Sasaki}},
  \bibinfo{journal}{JCAP} \textbf{\bibinfo{volume}{09}}, \bibinfo{pages}{012}
  (\bibinfo{year}{2013}), \eprint{1306.2901}.

\bibitem[{\citenamefont{Celoria et~al.}(2019)\citenamefont{Celoria, Comelli,
  Pilo, and Rollo}}]{Celoria:2019oiu}
\bibinfo{author}{\bibfnamefont{M.}~\bibnamefont{Celoria}},
  \bibinfo{author}{\bibfnamefont{D.}~\bibnamefont{Comelli}},
  \bibinfo{author}{\bibfnamefont{L.}~\bibnamefont{Pilo}}, \bibnamefont{and}
  \bibinfo{author}{\bibfnamefont{R.}~\bibnamefont{Rollo}},
  \bibinfo{journal}{JCAP} \textbf{\bibinfo{volume}{12}}, \bibinfo{pages}{018}
  (\bibinfo{year}{2019}), \eprint{1907.11784}.

\bibitem[{\citenamefont{Makino and Sasaki}(1991)}]{Makino:1991sg}
\bibinfo{author}{\bibfnamefont{N.}~\bibnamefont{Makino}} \bibnamefont{and}
  \bibinfo{author}{\bibfnamefont{M.}~\bibnamefont{Sasaki}},
  \bibinfo{journal}{Prog. Theor. Phys.} \textbf{\bibinfo{volume}{86}},
  \bibinfo{pages}{103} (\bibinfo{year}{1991}).

\bibitem[{\citenamefont{Weinberg}(2003)}]{Weinberg:2003sw}
\bibinfo{author}{\bibfnamefont{S.}~\bibnamefont{Weinberg}},
  \bibinfo{journal}{Phys. Rev. D} \textbf{\bibinfo{volume}{67}},
  \bibinfo{pages}{123504} (\bibinfo{year}{2003}), \eprint{astro-ph/0302326}.

\bibitem[{\citenamefont{Chiba and Yamaguchi}(2008)}]{Chiba:2008ia}
\bibinfo{author}{\bibfnamefont{T.}~\bibnamefont{Chiba}} \bibnamefont{and}
  \bibinfo{author}{\bibfnamefont{M.}~\bibnamefont{Yamaguchi}},
  \bibinfo{journal}{JCAP} \textbf{\bibinfo{volume}{10}}, \bibinfo{pages}{021}
  (\bibinfo{year}{2008}), \eprint{0807.4965}.

\bibitem[{\citenamefont{Sugiyama and Futamase}(2010)}]{Sugiyama:2010zz}
\bibinfo{author}{\bibfnamefont{N.}~\bibnamefont{Sugiyama}} \bibnamefont{and}
  \bibinfo{author}{\bibfnamefont{T.}~\bibnamefont{Futamase}},
  \bibinfo{journal}{Phys. Rev. D} \textbf{\bibinfo{volume}{81}},
  \bibinfo{pages}{023504} (\bibinfo{year}{2010}).

\bibitem[{\citenamefont{Kubota et~al.}(2012)\citenamefont{Kubota, Misumi,
  Naylor, and Okuda}}]{Kubota:2011re}
\bibinfo{author}{\bibfnamefont{T.}~\bibnamefont{Kubota}},
  \bibinfo{author}{\bibfnamefont{N.}~\bibnamefont{Misumi}},
  \bibinfo{author}{\bibfnamefont{W.}~\bibnamefont{Naylor}}, \bibnamefont{and}
  \bibinfo{author}{\bibfnamefont{N.}~\bibnamefont{Okuda}},
  \bibinfo{journal}{JCAP} \textbf{\bibinfo{volume}{02}}, \bibinfo{pages}{034}
  (\bibinfo{year}{2012}), \eprint{1112.5233}.

\bibitem[{\citenamefont{White}(2014)}]{White}
\bibinfo{author}{\bibfnamefont{J.}~\bibnamefont{White}}, \bibinfo{journal}{The
  Universe} \textbf{\bibinfo{volume}{2}}, \bibinfo{pages}{2}
  (\bibinfo{year}{2014}).


\bibitem[{\citenamefont{Skugoreva et~al.}(2014)\citenamefont{Skugoreva,
  Toporensky, and Vernov}}]{Skugoreva:2014gka}
\bibinfo{author}{\bibfnamefont{M.~A.} \bibnamefont{Skugoreva}},
  \bibinfo{author}{\bibfnamefont{A.~V.} \bibnamefont{Toporensky}},
  \bibnamefont{and} \bibinfo{author}{\bibfnamefont{S.~Y.}
  \bibnamefont{Vernov}}, \bibinfo{journal}{Phys. Rev. D}
  \textbf{\bibinfo{volume}{90}}, \bibinfo{pages}{064044}
  (\bibinfo{year}{2014}), \eprint{1404.6226}.

\bibitem[{\citenamefont{Pozdeeva et~al.}(2016)\citenamefont{Pozdeeva,
  Skugoreva, Toporensky, and Vernov}}]{Pozdeeva:2016cja}
\bibinfo{author}{\bibfnamefont{E.~O.} \bibnamefont{Pozdeeva}},
  \bibinfo{author}{\bibfnamefont{M.~A.} \bibnamefont{Skugoreva}},
  \bibinfo{author}{\bibfnamefont{A.~V.} \bibnamefont{Toporensky}},
  \bibnamefont{and} \bibinfo{author}{\bibfnamefont{S.~Y.}
  \bibnamefont{Vernov}}, \bibinfo{journal}{JCAP} \textbf{\bibinfo{volume}{12}},
  \bibinfo{pages}{006} (\bibinfo{year}{2016}), \eprint{1608.08214}.

\bibitem[{\citenamefont{Faraoni}(2007)}]{Faraoni:2007yn}
\bibinfo{author}{\bibfnamefont{V.}~\bibnamefont{Faraoni}},
  \bibinfo{journal}{Phys. Rev. D} \textbf{\bibinfo{volume}{75}},
  \bibinfo{pages}{067302} (\bibinfo{year}{2007}), \eprint{gr-qc/0703044}.

\bibitem[{\citenamefont{Vernov and Pozdeeva}(2021)}]{Vernov:2021hxo}
\bibinfo{author}{\bibfnamefont{S.}~\bibnamefont{Vernov}} \bibnamefont{and}
  \bibinfo{author}{\bibfnamefont{E.}~\bibnamefont{Pozdeeva}},
  \bibinfo{journal}{Universe} \textbf{\bibinfo{volume}{7}},
  \bibinfo{pages}{149} (\bibinfo{year}{2021}), \eprint{2104.11111}.

\bibitem[{\citenamefont{Pham et~al.}(2024)\citenamefont{Pham, Nguyen, Do, and
  Kao}}]{Pham:2024fub}
\bibinfo{author}{\bibfnamefont{T.~M.} \bibnamefont{Pham}},
  \bibinfo{author}{\bibfnamefont{D.~H.} \bibnamefont{Nguyen}},
  \bibinfo{author}{\bibfnamefont{T.~Q.} \bibnamefont{Do}}, \bibnamefont{and}
  \bibinfo{author}{\bibfnamefont{W.~F.} \bibnamefont{Kao}}
  (\bibinfo{year}{2024}), \eprint{2403.02623}.

\end{thebibliography}

\end{document}